\pdfoutput=1
\documentclass[sigconf, screen, nonacm]{acmart}

\setcopyright{none}
\renewcommand\footnotetextcopyrightpermission[1]{}

\usepackage{tikz}
\usetikzlibrary{arrows.meta,positioning,calc,fit,backgrounds,shapes.misc}
\usepackage{pgfplots}
\pgfplotsset{compat=1.18}
\usepackage{xcolor}
\usepackage{array}
\definecolor{cInk}{HTML}{262626}
\definecolor{cMute}{HTML}{7F7F7F}
\definecolor{cGrid}{HTML}{D9D9D9}
\definecolor{cBlue}{HTML}{23667A}
\definecolor{cOrange}{HTML}{923931}
\definecolor{cBlueL}{HTML}{DBEEF3}
\definecolor{cOrangeL}{HTML}{F2DCDA}
\definecolor{cGrayL}{HTML}{F2F2F2}

\newcommand{\secref}[1]{Section~\ref{#1}}
\newcommand{\figref}[1]{Figure~\ref{#1}}
\newcommand{\tabref}[1]{Table~\ref{#1}}
\newcommand{\sys}{Froq}
\newcommand{\ie}{\textit{i.e.}}


\begin{document}

\title[\texorpdfstring{\sys{}}{\sys}: Safe Speculative Execution on Live Quantum States for FTQC]{\texorpdfstring{\sys{}}{\sys}: Safe Speculative
  Execution\texorpdfstring{\\}{ }on Live Quantum States for FTQC}

\author{Yicheng Guang}
\affiliation{%
  \institution{University of Colorado Boulder}
  \city{Boulder}
  \state{Colorado}
  \country{USA}}
\email{yicheng.guang@colorado.edu}
\author{Yueqi Chen}
\affiliation{%
  \institution{University of Colorado Boulder}
  \city{Boulder}
  \state{Colorado}
  \country{USA}}
\email{yueqi.chen@colorado.edu}

\begin{abstract}
\looseness=-1
Reducing execution time is critical to fault-tolerant quantum computing (FTQC) as longer executions expose logical qubits to more opportunities for failure.
However, quantum processors could frequently stall due to slow classical error-correction decoders. 
Recent speculative decoding accelerates this process, but its speculation remains on the classical side, leaving quantum operations waiting for confirmed decoding results.
Extending speculative execution to the quantum side is fundamentally challenging. Unlike classical processor states, quantum states cannot be arbitrarily copied or checkpointed for rollback, making recovery from mispredictions difficult.
In this work, we propose \sys{} to enable safe speculative execution on live quantum states. 
\sys{} uses compiler-assisted recovery contracts to track how mispredictions propagate through quantum operations. 
It further employs a resource-aware runtime mechanism to reserve recovery workspace and dynamically balance the benefits of speculative execution against the expected costs of recovery.
We evaluate \sys{} on eight FTQC programs across 88 configurations, building on an existing speculative window decoder. 
\sys{} reduces quantum-side time by 9.81\% across all 88 configurations.
The reduction reaches 18.34\% across the 45 configurations allowing concurrent predictions, and peaks at 31.1\% at code distance \(d=21\), under limited resources.

\end{abstract}

\maketitle

\section{Introduction}\label{sec:intro}

\looseness=-1
Quantum computers are noisy.
Their qubits are easily disturbed by the environment and by imperfect control, so errors keep occurring throughout a computation.
To tolerate these errors, a quantum error correction (QEC) code encodes each logical qubit redundantly in many physical qubits and repeatedly measures checks on them~\cite{dennis2002,fowler2012}.
A classical decoder then analyzes the check results to locate and identify the errors, so that they can be corrected~\cite{dennis2002,terhal2015}.
By correcting errors continuously, QEC allows long quantum circuits to execute reliably and thus achieves the goal of fault-tolerant quantum computing (FTQC)~\cite{google2023}.

With QEC, reducing execution time is still critical to FTQC, as each additional round of checks gives errors another chance to slip past correction~\cite{google2023}.
A longer execution is more likely to fail, or needs a stronger code with more physical qubits to stay reliable.
Besides reliability, a shorter execution also frees the quantum computer sooner for the next program, improving the throughput and efficiency of the whole system.
However, quantum programs frequently stall to wait for the decoder's completion.
This is because some operations, such as the $T$ gate, depend on the decoding result to decide whether a correction must follow~\cite{litinski,swiper}.

Recent work shortens this wait by making the decoder faster.
SWIPER, for example, speculates inside the decoder.
It guesses intermediate results that one part of the decoding needs from another, so that dependent parts decode in parallel, and it redoes a part after a misprediction~\cite{swiper}.
However, this speculation stays on the classical side.
An instruction on the quantum processor that depends on a decoding result still waits until the decoder confirms that result.
As a result, even with SWIPER, over eight benchmark programs, the median of each program's mean wait per $T$ gate was 56.3 rounds of checks at code distance 13 and 214.7 at code distance 21.

\looseness=-1
A natural way to remove this stall is to also speculate on the quantum side, using a predicted decoding result and letting the dependent quantum operations proceed before the decoder finishes.
However, recovering from a misprediction is fundamentally harder on the quantum side.
A classical processor recovers from a mispredicted branch by discarding the speculative state that it keeps apart from the architectural state~\cite{smith1988}.
In contrast, a quantum state cannot be copied or checkpointed, so a quantum processor has no saved state to roll back to.
As a result, a misprediction will leave an error on the live qubits, and every later operation spreads this error further.
Instead of rolling back, the system must undo this error in place by applying extra quantum operations to the live qubits, and which operations are needed depends on how far the error has spread.

\looseness=-1
Such in-place recovery raises three challenges:
First, the system must know how to undo a misprediction wherever execution stops, while the recovery changes as execution advances (\textbf{C1}).
Second, a recovery needs extra qubits to run on.
If other operations occupy them when a misprediction is detected, the recovery must wait, and the program can deadlock (\textbf{C2}).
Third, a recovery costs time that depends on where execution stands, so speculation does not always pay off and the system must decide when to speculate (\textbf{C3}).

\looseness=-1
Prior work has started to run quantum operations ahead of their inputs, but none of it addresses all three challenges.
ARTERY predicts a measurement result that decides a later operation and pre-executes the predicted branch~\cite{artery}, but it does not target decoding results.
Pipelined execution lets operations start before their predecessors finish decoding and rolls part of the work back when a prediction fails~\cite{spequlation}.
However, it chooses how far to run ahead with fixed strategies, which do not account for a recovery whose cost depends on where execution stands.

In this work, we propose \sys{}\footnote{\sys{} is named after a leaping frog, which jumps ahead as speculative execution runs ahead of the decoder, with the g of frog replaced by q for quantum.} to enable safe speculative execution on live quantum states.
Our key insight is that, although a quantum state cannot be saved, the recovery from a misprediction can be prepared in advance for every point where execution may stop.
\sys{} realizes this insight with compiler-assisted recovery contracts, which address \textbf{C1}.
For each decoding result that may be predicted, the compiler tracks how a misprediction would spread through the later operations.
At every point where execution may stop, it prepares a recovery, a short sequence of quantum operations that undoes the spread error, and records how long the recovery takes and which extra qubits it needs.
The compiler examines every recovery by simulation and records all of them in a \textbf{recovery contract}.
A contract extends only as far as the compiler can validate a recovery, and execution never runs ahead past its end (\secref{sec:repair}).

To address \textbf{C2} and \textbf{C3}, \sys{} further employs a resource-aware runtime in the controller.
For \textbf{C2}, before execution runs ahead on a prediction, the runtime reserves the extra qubits of every recovery in the contract, so a recovery never waits for qubits and the program cannot deadlock.
For \textbf{C3}, the runtime decides, for each pending decoding result in each round, whether to wait for the decoder or to run ahead.
It weighs the time that running ahead saves against the expected cost of recovery, which depends on the chance of a misprediction and on the cost of recovering from where execution would stand when the decoder finishes.
The compiler solves this decision offline into tables, so the controller can make each decision using a single table lookup (\secref{sec:runtime}) without violating real-time constraints.
Multiple predictions may be outstanding concurrently, as long as the operations and extra qubits they affect do not overlap.

\looseness=-1
We evaluate \sys{} in simulation on eight FTQC programs, each at several code distances and numbers of decoding cores, for 88 configurations in total (\secref{sec:eval}).
We build \sys{} on top of SWIPER and compare it with SWIPER alone in quantum-side time, the time that the quantum processor takes to finish the program.
Every simulated run charges the controller's latency, so the comparison includes the time that \sys{}'s decisions take.
\sys{} reduced quantum-side time by 9.81\% across all 88 configurations.
The reduction reached 18.34\% across the 45 configurations that allow concurrent predictions, and it peaked at 31.1\% for a 5-qubit quantum phase estimation program at code distance 21.
At 38 configurations, \sys{} determined in training that speculation does not pay off, so it always waits and runs exactly as SWIPER alone does.

In summary, this work makes the following contributions:
\vspace{-5pt}
\begin{itemize}
  \item We propose \sys{} to enable safe speculative execution on live quantum states and thus reduce quantum execution time for fewer errors and higher throughput.
  \item \looseness=-1 We implement \sys{}'s compiler and runtime on SWIPER's simulator and the key blocks of its controller in RTL for an FPGA. The code is available at \url{https://github.com/a85tract/Froq}.
  \item We evaluate \sys{} on eight FTQC programs across a range of code distances and decoder resources, and show its effectiveness in reducing quantum-side time.
\end{itemize}

\section{Background and Problem Description}\label{sec:background}

\looseness=-1
This section explains why quantum programs stall on the decoder and what a recovery from misprediction involves.
We first describe how a fault-tolerant quantum computer produces decoding results and why programs stall waiting for them (\secref{sec:bg-value}).
We then describe how speculating on the quantum side could remove this stall and what a misprediction leaves behind (\secref{sec:bg-spec}).
Finally, we make \textbf{C1}, \textbf{C2}, and \textbf{C3} concrete on this machine (\secref{sec:problem}).

\ifdefined\FigArchDone \fi
\gdef\FigArchDone{}
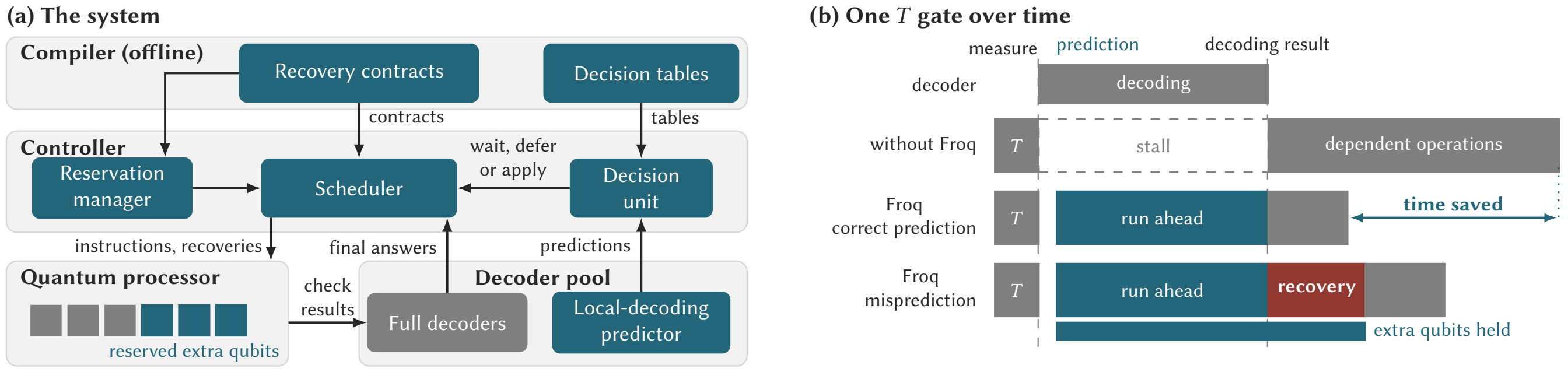
\begin{figure*}[t]
\centering
\begin{tikzpicture}[x=1mm,y=0.82mm,font=\sffamily\fontsize{7}{8.4}\selectfont,
  froq/.style={draw=cBlue,line width=0.6pt,fill=cBlue,rounded corners=0.8mm,align=center,text=white,inner sep=0.8mm,minimum height=6.5mm},
  base/.style={draw=cMute,line width=0.6pt,fill=cMute,rounded corners=0.8mm,align=center,text=white,inner sep=0.8mm,minimum height=6.5mm},
  region/.style={draw=cGrid,line width=0.45pt,fill=cGrayL,rounded corners=1.2mm},
  head/.style={font=\sffamily\fontsize{7.2}{8.4}\selectfont\bfseries,text=cInk,anchor=north west,inner sep=0pt},
  ttl/.style={font=\sffamily\fontsize{7.6}{9}\selectfont\bfseries,text=cInk,anchor=north west,inner sep=0pt},
  flow/.style={-{Latex[length=1.6mm,width=1.2mm]},line width=0.7pt,cInk},
  lab/.style={font=\sffamily\fontsize{6.4}{7.6}\selectfont,text=cInk,inner sep=0.4mm,align=center},
  patch/.style={draw=cMute,line width=0.4pt,fill=cMute,minimum size=3.4mm,inner sep=0pt},
  rpatch/.style={draw=cBlue,line width=0.7pt,fill=cBlue,minimum size=3.4mm,inner sep=0pt},
  bar/.style={draw=cMute,line width=0.5pt,fill=cMute,inner sep=0pt,font=\sffamily\fontsize{6.4}{7.6}\selectfont,text=white},
  spec/.style={bar,draw=cBlue,fill=cBlue},
  guide/.style={cMute,dashed,line width=0.4pt}]
  \node[ttl] at (0,49.8) {(a) The system};
  \draw[region] (0,35.5) rectangle (84,45.5); \node[head] at (1.5,44.6) {Compiler (offline)};
  \node[froq,minimum width=27mm] (rc) at (40,40.5) {Recovery contracts};
  \node[froq,minimum width=22mm] (dt) at (72,40.5) {Decision tables};
  \draw[region] (0,18.5) rectangle (84,32.5); \node[head] at (1.5,31.5) {Controller};
  \node[froq,minimum width=18mm] (rm) at (12,24.6) {Reservation\\manager};
  \node[froq,minimum width=22mm] (sc) at (40,24.6) {Scheduler};
  \node[froq,minimum width=16mm] (du) at (72,24.6) {Decision\\unit};
  \draw[region] (0,0) rectangle (32,14.5); \node[head] at (1.5,13.5) {Quantum processor};
  \foreach \i in {0,...,5} { \node[patch] at ({4.5+4.2*\i},6.3) {}; }
  \foreach \i in {3,4,5} { \node[rpatch] at ({4.5+4.2*\i},6.3) {}; }
  \node[lab,text=cBlue] at (21.3,1.9) {reserved extra qubits};
  \draw[region] (40,0) rectangle (84,14.5); \node[head] at (53,13.5) {Decoder pool};
  \node[base,minimum width=18mm] (fd) at (50,6) {Full decoders};
  \node[froq,minimum width=20mm] (pr) at (72,6) {Local-decoding\\predictor};
  \draw[flow] (dt.south) -- node[lab,right=0.6mm,pos=0.25] {tables} (du.north);
  \draw[flow] (rc.south -| sc.north) -- node[lab,right=0.6mm,pos=0.25] {contracts} (sc.north);
  \draw[flow] (rc.west) -| (rm.north -| 18,0);
  \draw[flow] (du.west) -- node[lab,above=0.4mm] {wait, defer\\or apply} (sc.east);
  \draw[flow] (rm.east) -- (sc.west);
  \draw[flow] (sc.south -| 30,0) -- (30,14.5);
  \node[lab,anchor=east] at (29.5,16.5) {instructions, recoveries};
  \draw[flow] (32,6) -- node[lab,above=0.3mm] {check\\results} (fd.west);
  \draw[flow] (fd.north) -- (sc.south -| fd.north);
  \node[lab,anchor=east] at (49.3,16.5) {final answers};
  \draw[flow] (pr.north) -- (du.south);
  \node[lab,anchor=east] at (71.3,16.5) {predictions};
  \begin{scope}[xshift=91mm]
  \node[ttl] at (0,49.8) {(b) One $T$ gate over time};
  \draw[guide] (26,42.5) -- (26,2.5); \draw[guide] (52,42.5) -- (52,2.5);
  \node[lab,anchor=south east] at (26.4,42.5) {measure};
  \node[lab,anchor=south west,text=cBlue] at (27.6,42.5) {prediction};
  \node[lab,anchor=south] at (52,42.5) {decoding result};
  \node[lab,anchor=east] at (19.5,39) {decoder};
  \node[bar,minimum height=4.4mm,minimum width=26mm,anchor=west] at (26,39) {decoding};
  \node[lab,anchor=east] at (19.5,30.5) {without \sys{}};
  \node[bar,minimum height=6mm,minimum width=5mm,anchor=west] at (21,30.5) {$T$};
  \node[bar,fill=white,dashed,text=cMute,minimum height=6mm,minimum width=26mm,anchor=west] at (26,30.5) {stall};
  \node[bar,minimum height=6mm,minimum width=33mm,anchor=west] at (52,30.5) {dependent operations};
  \node[lab,anchor=east] at (19.5,20.5) {\sys{}\\correct prediction};
  \node[bar,minimum height=6mm,minimum width=5mm,anchor=west] at (21,20.5) {$T$};
  \node[spec,minimum height=6mm,minimum width=24mm,anchor=west] at (28,20.5) {run ahead};
  \node[bar,minimum height=6mm,minimum width=9mm,anchor=west] at (52,20.5) {};
  \draw[cBlue,dotted,line width=0.6pt] (85,27.5) -- (85,20.5);
  \draw[cBlue,{Latex[length=1.4mm,width=1.1mm]}-{Latex[length=1.4mm,width=1.1mm]},line width=0.7pt] (61.4,20.5) -- (84.8,20.5);
  \node[lab,text=cBlue,font=\sffamily\fontsize{6.6}{7.6}\selectfont\bfseries,anchor=south] at (73.1,20.9) {time saved};
  \node[lab,anchor=east] at (19.5,10.5) {\sys{}\\misprediction};
  \node[bar,minimum height=6mm,minimum width=5mm,anchor=west] at (21,10.5) {$T$};
  \node[spec,minimum height=6mm,minimum width=24mm,anchor=west] at (28,10.5) {run ahead};
  \node[bar,draw=cOrange,fill=cOrange,text=white,font=\sffamily\fontsize{6.4}{7.6}\selectfont\bfseries,minimum height=6mm,minimum width=11mm,anchor=west] at (52,10.5) {recovery};
  \node[bar,minimum height=6mm,minimum width=9mm,anchor=west] at (63,10.5) {};
  \node[spec,minimum height=2mm,minimum width=35mm,anchor=west] at (28,4.8) {};
  \node[lab,anchor=west,text=cBlue] at (63.6,4.8) {extra qubits held};
  \end{scope}
\end{tikzpicture}
\caption{(a) The components that \sys{} adds to an FTQC system.
  (b) The effect of \sys{} on one $T$ gate.}
\label{fig:arch}
\end{figure*}

\subsection{Why programs stall on the decoder}\label{sec:bg-value}

\looseness=-1
A quantum error correction code such as the surface code protects each logical qubit with a square of physical qubits called a patch~\cite{dennis2002,fowler2012,google2023}.
The code distance $d$ of a patch sets how many physical errors the patch can correct.
In every syndrome round, or simply round, the hardware measures the parity checks of every patch, and a classical decoder uses their outcomes to infer which errors occurred.

\looseness=-1
To perform a logical operation, lattice surgery merges patches through free cells of the layout's grid and measures them jointly~\cite{horsman2012,litinski}.
The decoder processes the check outcomes as a stream of rounds and splits this stream into windows, each of which depends on the decoding results of its neighbors at their shared edges~\cite{dennis2002,terhal2015,skoric2023,tan2023}.
Sliding-window decoding decodes a window only after the windows it depends on are decoded.
In contrast, SWIPER's speculative window decoding predicts the result at a shared edge, decodes dependent windows in parallel, and re-decodes a window after a misprediction~\cite{swiper}.
SWIPER shortens application runtimes by 40\% on average compared with prior parallel window decoders~\cite{swiper}.
However, neither sliding-window decoding nor SWIPER lets the program run ahead of the decoder, so an operation that needs a decoding result still waits for it.

Such operations include the $S$ correction of a $T$ gate and the measurement-conditioned $CZ$ of an uncomputation~\cite{gidney2018}.
This paper focuses on the $T$ gate.
We call the lattice-surgery template that implements one logical operation a gadget, and the $T$ gate runs as a teleportation gadget~\cite{gottesman1999,bravyi2005,litinski2019msd}.
This gadget measures a $ZZ$ parity, and the decoded outcome of this measurement decides whether a corrective phase gate $S$ must follow.
The measurement finishes before the decoder produces this outcome, and we call a correction that waits for such an outcome a \textbf{pending correction}.

A pending correction does not always stall the program.
If the correction is a Pauli operator (\ie, $X$, $Y$, or $Z$), software records it in a Pauli frame instead of applying it to the qubits~\cite{knill2005,pauliframes}.
However, $S$ is a quarter-turn of phase, which is not a Pauli operator, so the Pauli frame cannot absorb it, and the operations that depend on $S$ must wait for the decoder.
Auto-corrected rotations~\cite{litinski} avoid this wait by moving the conditional correction onto an extra patch, which waits for the decoder while the next operation proceeds.
The cost is that the extra patch occupies cells of the layout until the decoder returns the outcome.
When decoding is slow, several such patches can be pending at once, so the layout either needs more physical qubits or leaves fewer free cells for lattice surgery, which can delay other operations.
Thus, a solution that removes this wait without adding patches must let the operations that depend on $S$ proceed before the outcome is decoded, and must correct the qubits when the outcome differs from what these operations assumed.

\subsection{Speculating on live quantum states}\label{sec:bg-spec}
Speculation lets the operations that depend on a pending correction proceed before the outcome is decoded.
Independent of the decoder, each such operation has three options.
First, it can wait for the decoded outcome, as SWIPER alone does.
Second, it can defer the correction, betting that no $S$ is needed, and run without it.
A Clifford frame, which unlike the Pauli frame can hold corrections such as $S$, records the deferred $S$~\cite{chamberland2018}.
Third, it can apply the predicted correction, betting on the prediction.
Deferring and applying both run ahead of the decoder, so both speculate.

However, speculating on live quantum states makes recovery from a misprediction hard.
A classical processor can discard its speculative state, but a quantum state cannot be copied or checkpointed, so there is no saved state to roll back to.
When the decoded outcome contradicts the bet, the speculated operations have already acted on the live qubits and left an error on them, which we call the \textbf{residual}.
The system must therefore undo the residual in place, by applying extra quantum operations to the live qubits.

Which operations undo the residual depends on how far it has spread, because a residual spreads through every later operation on its qubits.
For example, a mispredicted $S$ on qubit $q_0$ leaves an unwanted quarter-turn on $q_0$.
A later CNOT from $q_1$ to $q_0$ spreads this rotation to both qubits and turns it into a joint quarter-turn about $Z_0Z_1$.
A recovery at this point must undo exactly that joint rotation, not the original quarter-turn on $q_0$.
By analogy with the light cone in physics, which bounds the events that a cause can influence, we call the operations and measurement outcomes that a misprediction can reach in this way its \textbf{light cone}.
When the decoder reveals a misprediction, execution has progressed to some point inside the light cone.
We call each point at which execution can pause for a recovery a \textbf{boundary}.
At each boundary, the system undoes the residual by running a short sequence of quantum operations called a \textbf{recovery script}.
The script runs on extra qubits, and the free cells that hold them form its \textbf{recovery workspace}.

\subsection{Challenges of in-place recovery}\label{sec:problem}

In-place recovery on live quantum states raises three challenges.

\textbf{C1: A checked recovery script for every boundary.}
The residual differs from boundary to boundary, so the recovery script does too.
Lattice-surgery gadgets also contain measurements, and a residual that reaches one can flip or relabel its outcome, so a script must also correct the recorded outcomes that later operations read.
Deriving and checking a script only after the decoder reveals a misprediction would delay the recovery, so a design must prepare a checked script for every boundary in advance.

\looseness=-1
\textbf{C2: A free workspace whenever a recovery needs it.}
A recovery script can start only when its workspace is free, and speculation can make the workspace wait on the recovery itself.
For example, suppose a cell of the workspace holds a patch that only the next instruction on that cell discards, and that instruction reads the qubit that the pending correction targets.
After a misprediction, the recovery waits for the cell, the cell waits for that instruction, and the instruction waits for the correction, which settles only through the recovery.
Without speculation no recovery exists and this cycle cannot form, but with speculation a single speculative correction suffices to deadlock the program.

\textbf{C3: A decision for each pending correction in each round.}
Speculation pays only if the rounds it saves outweigh the expected cost of recovery, namely the probability of a misprediction times the cost of recovering.
This cost depends on the boundary where execution stands when the decode ends, and on the decoder load, because the results of a recovery must also be decoded.
The misprediction probability, the decode time and the decoder load change across corrections and rounds, so a design must decide for each pending correction in each round whether to speculate.

\vspace{-4pt}

\ifdefined\FigArchDone \fi
\gdef\FigArchDone{}
\begin{figure*}[t]
\centering
\begin{tikzpicture}[x=1mm,y=0.82mm,font=\sffamily\fontsize{7}{8.4}\selectfont,
  froq/.style={draw=cBlue,line width=0.6pt,fill=cBlue,rounded corners=0.8mm,align=center,text=white,inner sep=0.8mm,minimum height=6.5mm},
  base/.style={draw=cMute,line width=0.6pt,fill=cMute,rounded corners=0.8mm,align=center,text=white,inner sep=0.8mm,minimum height=6.5mm},
  region/.style={draw=cGrid,line width=0.45pt,fill=cGrayL,rounded corners=1.2mm},
  head/.style={font=\sffamily\fontsize{7.2}{8.4}\selectfont\bfseries,text=cInk,anchor=north west,inner sep=0pt},
  ttl/.style={font=\sffamily\fontsize{7.6}{9}\selectfont\bfseries,text=cInk,anchor=north west,inner sep=0pt},
  flow/.style={-{Latex[length=1.6mm,width=1.2mm]},line width=0.7pt,cInk},
  lab/.style={font=\sffamily\fontsize{6.4}{7.6}\selectfont,text=cInk,inner sep=0.4mm,align=center},
  patch/.style={draw=cMute,line width=0.4pt,fill=cMute,minimum size=3.4mm,inner sep=0pt},
  rpatch/.style={draw=cBlue,line width=0.7pt,fill=cBlue,minimum size=3.4mm,inner sep=0pt},
  bar/.style={draw=cMute,line width=0.5pt,fill=cMute,inner sep=0pt,font=\sffamily\fontsize{6.4}{7.6}\selectfont,text=white},
  spec/.style={bar,draw=cBlue,fill=cBlue},
  guide/.style={cMute,dashed,line width=0.4pt}]
  \node[ttl] at (0,49.8) {(a) The system};
  \draw[region] (0,35.5) rectangle (84,45.5); \node[head] at (1.5,44.6) {Compiler (offline)};
  \node[froq,minimum width=27mm] (rc) at (40,40.5) {Recovery contracts};
  \node[froq,minimum width=22mm] (dt) at (72,40.5) {Decision tables};
  \draw[region] (0,18.5) rectangle (84,32.5); \node[head] at (1.5,31.5) {Controller};
  \node[froq,minimum width=18mm] (rm) at (12,24.6) {Reservation\\manager};
  \node[froq,minimum width=22mm] (sc) at (40,24.6) {Scheduler};
  \node[froq,minimum width=16mm] (du) at (72,24.6) {Decision\\unit};
  \draw[region] (0,0) rectangle (32,14.5); \node[head] at (1.5,13.5) {Quantum processor};
  \foreach \i in {0,...,5} { \node[patch] at ({4.5+4.2*\i},6.3) {}; }
  \foreach \i in {3,4,5} { \node[rpatch] at ({4.5+4.2*\i},6.3) {}; }
  \node[lab,text=cBlue] at (21.3,1.9) {reserved extra qubits};
  \draw[region] (40,0) rectangle (84,14.5); \node[head] at (53,13.5) {Decoder pool};
  \node[base,minimum width=18mm] (fd) at (50,6) {Full decoders};
  \node[froq,minimum width=20mm] (pr) at (72,6) {Local-decoding\\predictor};
  \draw[flow] (dt.south) -- node[lab,right=0.6mm,pos=0.25] {tables} (du.north);
  \draw[flow] (rc.south -| sc.north) -- node[lab,right=0.6mm,pos=0.25] {contracts} (sc.north);
  \draw[flow] (rc.west) -| (rm.north -| 18,0);
  \draw[flow] (du.west) -- node[lab,above=0.4mm] {wait, defer\\or apply} (sc.east);
  \draw[flow] (rm.east) -- (sc.west);
  \draw[flow] (sc.south -| 30,0) -- (30,14.5);
  \node[lab,anchor=east] at (29.5,16.5) {instructions, recoveries};
  \draw[flow] (32,6) -- node[lab,above=0.3mm] {check\\results} (fd.west);
  \draw[flow] (fd.north) -- (sc.south -| fd.north);
  \node[lab,anchor=east] at (49.3,16.5) {final answers};
  \draw[flow] (pr.north) -- (du.south);
  \node[lab,anchor=east] at (71.3,16.5) {predictions};
  \begin{scope}[xshift=91mm]
  \node[ttl] at (0,49.8) {(b) One $T$ gate over time};
  \draw[guide] (26,42.5) -- (26,2.5); \draw[guide] (52,42.5) -- (52,2.5);
  \node[lab,anchor=south east] at (26.4,42.5) {measure};
  \node[lab,anchor=south west,text=cBlue] at (27.6,42.5) {prediction};
  \node[lab,anchor=south] at (52,42.5) {decoding result};
  \node[lab,anchor=east] at (19.5,39) {decoder};
  \node[bar,minimum height=4.4mm,minimum width=26mm,anchor=west] at (26,39) {decoding};
  \node[lab,anchor=east] at (19.5,30.5) {without \sys{}};
  \node[bar,minimum height=6mm,minimum width=5mm,anchor=west] at (21,30.5) {$T$};
  \node[bar,fill=white,dashed,text=cMute,minimum height=6mm,minimum width=26mm,anchor=west] at (26,30.5) {stall};
  \node[bar,minimum height=6mm,minimum width=33mm,anchor=west] at (52,30.5) {dependent operations};
  \node[lab,anchor=east] at (19.5,20.5) {\sys{}\\correct prediction};
  \node[bar,minimum height=6mm,minimum width=5mm,anchor=west] at (21,20.5) {$T$};
  \node[spec,minimum height=6mm,minimum width=24mm,anchor=west] at (28,20.5) {run ahead};
  \node[bar,minimum height=6mm,minimum width=9mm,anchor=west] at (52,20.5) {};
  \draw[cBlue,dotted,line width=0.6pt] (85,27.5) -- (85,20.5);
  \draw[cBlue,{Latex[length=1.4mm,width=1.1mm]}-{Latex[length=1.4mm,width=1.1mm]},line width=0.7pt] (61.4,20.5) -- (84.8,20.5);
  \node[lab,text=cBlue,font=\sffamily\fontsize{6.6}{7.6}\selectfont\bfseries,anchor=south] at (73.1,20.9) {time saved};
  \node[lab,anchor=east] at (19.5,10.5) {\sys{}\\misprediction};
  \node[bar,minimum height=6mm,minimum width=5mm,anchor=west] at (21,10.5) {$T$};
  \node[spec,minimum height=6mm,minimum width=24mm,anchor=west] at (28,10.5) {run ahead};
  \node[bar,draw=cOrange,fill=cOrange,text=white,font=\sffamily\fontsize{6.4}{7.6}\selectfont\bfseries,minimum height=6mm,minimum width=11mm,anchor=west] at (52,10.5) {recovery};
  \node[bar,minimum height=6mm,minimum width=9mm,anchor=west] at (63,10.5) {};
  \node[spec,minimum height=2mm,minimum width=35mm,anchor=west] at (28,4.8) {};
  \node[lab,anchor=west,text=cBlue] at (63.6,4.8) {extra qubits held};
  \end{scope}
\end{tikzpicture}
\caption{(a) The components that \sys{} adds to an FTQC system.
  (b) The effect of \sys{} on one $T$ gate.}
\label{fig:arch}
\end{figure*}

\section{\sys{} Overview}\label{sec:overview}

\figref{fig:arch} shows how \sys{} speculates safely on live quantum states in an FTQC system.
In \figref{fig:arch}a, the compiler builds recovery contracts offline, which prepare a checked recovery script for every boundary that a misprediction can reach (\textbf{C1}).
The controller runs the resource-aware runtime online, which holds the workspace of these scripts until the correction settles (\textbf{C2}) and decides in each round whether to speculate (\textbf{C3}), and the predictor runs on idle cores of the decoder pool.
\figref{fig:arch}b shows the effect on one $T$ gate.
Without \sys{}, the dependent operations stall until the decoding result arrives, whereas \sys{} runs them ahead on the prediction and, after a misprediction, runs the prepared recovery in place.

In the following, we first describe the predictions that \sys{} uses, and the roles of the compiler, the controller and the decoder.
\secref{sec:repair} and \secref{sec:runtime} will provide more technical details.

\subsection{Predictions and the predictor}\label{sec:client}

To apply a correction early, \sys{} needs a prediction.
For each pending correction, a predictor publishes three items: a predicted outcome $\hat m$, its confidence $c$, and the decode whose final answer will confirm or contradict it.
\sys{} makes no assumption about how the prediction is made and allows a prediction to arrive late or never.
Any source that supplies these three items can drive \sys{}.

\sys{} speculates on the conditional $S$ of a $T$ gadget, which its decoded $ZZ$ parity selects.
The recovery contracts undo residuals from the recovery library, which holds single quarter-turns on at most two qubits.
In contrast, the measurement-conditioned $CZ$ of the adders decomposes into three quarter-turns, about $Z_a$, $Z_b$ and $Z_aZ_b$.
A misprediction there would leave a residual outside the library, so \sys{} lets the $CZ$ wait for its decode.

\looseness=-1
To predict the $T$ gadget's parity, \sys{} uses a local-decoding predictor.
It computes the parity by locally decoding the checks within $\lfloor d/2\rfloor$ lines of the seam where the two patches merge.
This local decode runs in chained windows that start from the full decoder's latest correction record for the patch.
The predictor reports as its confidence the lower $95\%$ bound of the accuracy measured for that gadget kind on the patch, binned by the pending correction's age at $d$, $3d$, $6d$ and $12d$ rounds.

\subsection{Compiler, controller and decoder}\label{sec:roles}

\looseness=-1
Before the program runs, the compiler prepares what the controller needs to act on each prediction.
For each correction that may be predicted, it traces the light cone through the gadget templates and generates a recovery contract, which lists the steps between boundaries and ends at a fence before the first step whose recovery it cannot validate.
It also precomputes the wait, defer or apply decision into tables, so that the controller decides with one table lookup.

The controller manages each pending correction from its prediction until its final answer.
When the predictor publishes a prediction, the reservation manager reserves the recovery workspace of every script in the correction's contract before the correction speculates.
In every round, the decision unit looks up, for each speculative correction, the table row indexed by the prediction's confidence $c$ and the correction's age, and returns wait, defer or apply.
The scheduler then issues the steps that the decision admits.
When the decode named in the prediction returns its final answer, an answer that confirms the predicted outcome $\hat m$ removes the dependency, and one that contradicts $\hat m$ causes the scheduler to insert a recovery script into the program.
\sys{} may speculate on several pending corrections at once if their light cones and workspaces do not overlap, and we call these concurrent predictions.

Finally, the full decoders and the local-decoding predictor run on one shared pool of cores, whose size we call the decoder budget.
\sys{} does not reserve cores for the predictor, because every pending correction still needs a full decode to confirm it.
When cores are scarce, cores reserved for the predictor would delay these confirmations, whereas idle cores cost full decoding nothing.
Full decoding has priority and processes the oldest pending windows whose predecessors are decoded.
The predictor runs only on idle cores, and if no core is idle, a full decode preempts the most recently started local decode.
Applying a predicted correction executes a recovery script early, and the results of this script must also be decoded.
Therefore, the runtime applies a predicted correction only if the idle cores, after subtracting those used by pending and running applications, suffice to decode the results of its script.

\section{Compiler-Assisted Recovery Contracts}\label{sec:repair}

\looseness=-1
To address \textbf{C1}, \sys{}'s compiler records how a misprediction spreads and how to undo it wherever execution stops, in a recovery contract for every predicted correction.
This section first defines the residuals that a misprediction leaves and the recovery library that undoes them, then walks the light cone, places and checks a recovery script at every stopping boundary, and closes with a worked contract.

\subsection{Mispredictions and the recovery library}\label{sec:residual}

A misprediction of a $T$ gadget's correction leaves $S$ or $S^\dagger$ on the corrected wire.
At each boundary, the residual relative to definitive execution, the run that waits for every decode, is $E=PQ$, up to global phase.
Here $P$ is a Pauli frame, possibly dependent on recorded outcomes, and $Q$ belongs to the recovery library
\[
  \mathrm{Lib}(A)=\{I\}\cup\{e^{\pm i\pi P'/4}: P'\text{ a Pauli product on }A,\ \mathrm{wt}(P')\le 2\},
\]
\looseness=-1
where $A$ contains the affected wires.
Recovery respects operator order, with $E^\dagger=Q^\dagger P^\dagger=(Q^\dagger P^\dagger Q)Q^\dagger$.
The device executes $Q^\dagger$ and then the frame $Q^\dagger P^\dagger Q$, which is Pauli because every library element is Clifford.
A residual with no quarter-turn needs only bookkeeping.

\subsection{Walking the light cone}\label{sec:walk}

To track how this residual propagates, the compiler walks the light cone of the predicted correction through the program.
It first binds each logical operation to its instructions, patches and timing, with its quantum, classical and physical dependencies.
The walk then starts at the predicted correction with state $(E,A,D)$, the residual, the affected wires and the symbolic classical dependencies.
At first, $A$ holds the corrected wire and $D$ the predicted result.
The frontier follows affected wires, result consumers and outputs, and a wire joins at the gadget it first touches.
For every outcome record $r$, a step that the walk admits (\secref{sec:steps}) satisfies $K_rE=E'_rK_{f(r)}$, where $K_r$ is the operation that the step applies when its outcomes are $r$, and $f$ relabels results.
The identity updates $E$, $A$ and $D$.
For example, in a lattice-surgery CNOT from control $c$ to target $t$, $K_r=X_t^{a(r)}Z_c^{b(r)}\,\mathrm{CNOT}$ up to a scalar, with $a(r)$ and $b(r)$ parities of its recorded outcomes.
A quarter-turn entering on the target therefore leaves on both wires,
\[
  K_r\,e^{i\pi sZ_t/4}=(Z_cZ_t)^{a(r)}\,e^{i\pi sZ_cZ_t/4}\,K_r,\qquad s=\pm1,
\]
up to phase.
The quarter-turn is the same for every outcome, its frame depends on the recorded outcomes, and $f$ is the identity.
A missing identity, or a canonical output outside $\mathrm{Lib}$, ends the walk at a fence.
A destructive measurement is a fence too when the compiler cannot represent its reduced instrument, the quantum operation on the remaining qubits with its outcomes.
So are classical consumers of $D$, outputs written from $A$ and the publication of $D$.

\subsection{Steps, recovery scripts and settlement}\label{sec:steps}

\looseness=-1
The walk splits the light cone into steps, and the contract places a recovery script at the stopping boundary after each step.

\looseness=-1
\textbf{Steps.}
A pending correction admits at most 16 boundaries, and its steps last 1 to 64 syndrome rounds.
An admitted step completes when its last instruction ends.

\textbf{Recovery scripts.}
Each recovery script implements $Q^\dagger$ as the shortest sequence of checked templates, followed by one frame Pauli, and the router places it on the program's layout.
The templates are $S$, patched CNOT, Hadamard and a direct quarter-turn about $X$, which is the teleported-$S$ gadget on its other boundary and is checked by the same rule.
The recovery from a quarter-turn about $X$ is therefore one gadget, whose script lasts 21, 27 and 33 rounds at $d=13$, $17$ and $21$.
For a quarter-turn about $Y$, the recovery uses $S$, the quarter-turn about $X$, and $S$ only when this sequence is cheaper than decomposing with Hadamard gates, which held only at $d=13$.
A placed script's length is its critical path in rounds, its recovery workspace is the set of auxiliary cells it occupies apart from the program cells it acts on, and its volume counts its patch-rounds.
The contract also gives the rounds until the script's results are decoded and its decoding demand.
The runtime reads these numbers to reserve the workspace and to price the recovery (\secref{sec:runtime}).

\looseness=-1
\textbf{Settlement.}
When the final answer reveals a misprediction, the recovery starts only after the admitted steps before its boundary have ended, even when the answer arrives in flight.
The scheduler never cancels a partially executed measurement, and runs the shortest script between the current boundary and the farthest one reachable within the contract, so execution may continue past the misprediction.
The pending correction stays held until its recovery's results are decoded.
The decision's recovery price (\secref{sec:prices}) charges the script where an action leaves the correction.

\subsection{Checking every recovery script}\label{sec:cert}

The compiler checks every recovery script by simulation at every stopping boundary.
It first checks the templates' Kraus operators and byproducts exactly on at most 8 qubits, the propagation identities as canonical $(P,Q)$ pairs, $Q^\dagger$ and its frames for every library element, and each contract's members, order, states, event coverage, fences, placement, timing and cost.
Its region check then composes each admitted prefix with its placed recovery script in both directions of the misprediction, by direct comparison of the complete input-output quantum maps within 6 wires and 12 public bits, the outcomes that the program reads.
At run time, a monitor invalidates execution outside the admitted steps on every wire or result that a mispredicted correction may have reached.

\looseness=-1
\textbf{Correctness of a recovery contract.}
Fix either direction of the misprediction.
Let $a_1\ldots a_k$ be the admitted steps after the predicted correction, whose residual is $E_0=S^{\hat m-m}$.
Let $o$ be the public outcome record, with the private outcomes internal to a gadget summed after their frames are incorporated, and write $E_i=P_iQ_i$ for the residual after step $i$, which depends on the public history.
\begin{proposition}\label{prop:equiv}
\looseness=-1
Suppose every outcome of every admitted step satisfies its checked propagation identity $K_rE_{i-1}=E_iK_{f_i(r)}$
(allowing branch phases), that these relabelings induce a bijection $f$ on public records, that $Q_k\in\mathrm{Lib}$, and
that the recovery script implements $Q_k^\dagger$ followed by the frame $Q_k^\dagger P_k^\dagger Q_k$. Then the
speculative instrument with its recovery equals the definitive one, $\mathcal I^{\mathrm{spec}}_o(\rho)=\mathcal I^{\mathrm{def}}_{f(o)}(\rho)$,
for every public record $o$ and input $\rho$, after the recovery script and its frame.
\end{proposition}
\noindent\emph{Proof sketch.}
The checked entry gadget fixes the initial residual, and induction over the propagation identities gives the map before recovery $E_k(o)\,\mathcal I^{\mathrm{def}}_{f(o)}(\rho)\,E_k(o)^\dagger$.
The script and the frame implement $E_k(o)^\dagger=(Q_k^\dagger P_k^\dagger Q_k)\,Q_k^\dagger$, and the frame is Pauli because $Q_k$ is Clifford.
Branch phases cancel in the completely positive maps.
$\square$

Under ideal instruments, the proposition makes the recovery from any misprediction exact for the supported residuals and checked templates.
On the eight programs' evaluated schedules at $d=13$, $17$ and $21$, all 433,533 recovery regions passed by direct comparison.
Concurrent predictions stay apart, so the proposition applies to each one separately.
A step is admitted only if its light cone and classical dependencies are disjoint from every other pending correction's, no cell of its workspace is reserved for another correction, and no member waits on another correction's fence, pending cone work or reserved cell.

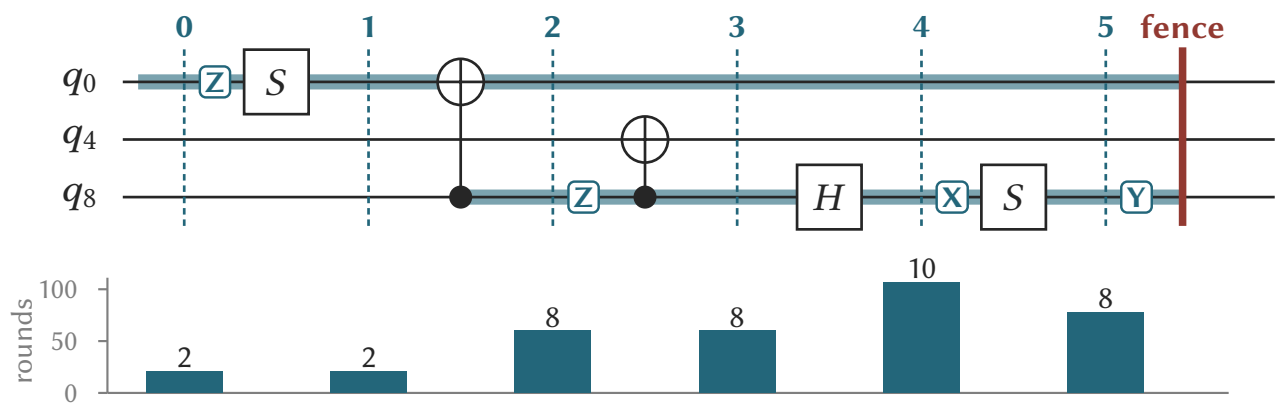
\begin{figure}[t]
\centering
\resizebox{\columnwidth}{!}{%
\begin{tikzpicture}[x=1mm,y=0.75mm,font=\sffamily\fontsize{7}{8.4}\selectfont,
  gate/.style={draw=cInk,line width=0.5pt,fill=white,minimum size=4.2mm,inner sep=0pt,text=cInk},
  res/.style={cBlue!60,line width=2.8pt},
  axis mark/.style={fill=white,draw=cBlue,line width=0.45pt,rounded corners=0.4mm,inner sep=0.35mm,text=cBlue,font=\sffamily\bfseries\fontsize{5.6}{6}\selectfont},
  bnd/.style={draw=cBlue,line width=0.5pt,dash pattern=on 1.2pt off 0.9pt}]
  \draw[res] (9,27) -- (77,27);
  \draw[res] (30,17) -- (77,17);
  \foreach \y/\q in {27/0,22/4,17/8} {
    \draw[cInk,line width=0.5pt] (8,\y) -- (83,\y);
    \node[anchor=east,text=cInk] at (7.6,\y) {$q_{\q}$};
  }
  \foreach \k/\x in {0/12,1/24,2/36,3/48,4/60,5/72} {
    \draw[bnd] (\x,14.5) -- (\x,30);
    \node[text=cBlue,font=\sffamily\bfseries\fontsize{6.4}{7}\selectfont] at (\x,31.8) {\k};
  }
  \node[gate] at (18,27) {$S$};
  \foreach \x/\y in {30/27,42/22} {
    \draw[cInk,line width=0.5pt] (\x,17) -- (\x,\y);
    \fill[cInk] (\x,17) circle (0.75mm);
    \draw[cInk,line width=0.5pt,fill=white] (\x,\y) circle (1.5mm);
    \draw[cInk,line width=0.5pt] (\x-1.5,\y) -- (\x+1.5,\y) ([yshift=-1.5mm]\x,\y) -- ([yshift=1.5mm]\x,\y);
  }
  \node[gate] at (54,17) {$H$};
  \node[gate] at (66,17) {$S$};
  \foreach \x/\y/\m in {14/27/Z,38/17/Z,62/17/X,74/17/Y} { \node[axis mark] at (\x,\y) {\m}; }
  \draw[cOrange,line width=1.4pt] (77,14.5) -- (77,30);
  \node[text=cOrange,font=\sffamily\bfseries\fontsize{6.4}{7}\selectfont] at (77,31.8) {fence};
  \draw[cMute,line width=0.4pt] (7,0) -- (7,10) (7,0) -- (80,0);
  \foreach \v in {0,50,100} {
    \draw[cMute,line width=0.4pt] (6.4,0.09*\v) -- (7,0.09*\v);
    \node[anchor=east,text=cMute,font=\sffamily\fontsize{5.4}{6}\selectfont] at (6.2,0.09*\v) {\v};
  }
  \node[rotate=90,text=cMute,font=\sffamily\fontsize{5.6}{6}\selectfont] at (1.4,4.5) {rounds};
  \foreach \x/\t/\c in {12/21/2,24/21/2,36/60/8,48/60/8,60/107/10,72/78/8} {
    \fill[cBlue] (\x-2.5,0) rectangle (\x+2.5,0.09*\t);
    \node[anchor=south,text=cInk,font=\sffamily\fontsize{6}{7}\selectfont,inner sep=0.5pt] at (\x,0.09*\t) {\c};
  }
\end{tikzpicture}}
\caption{One contract of Adder-4 at $d=13$ (\secref{sec:example}). Shading marks the wires that the residual occupies,
  with its axis boxed, and each bar is the recovery script's length at that boundary, with its cells above it.}
\label{fig:contract}
\end{figure}

\subsection{A worked contract}\label{sec:example}

\looseness=-1
\figref{fig:contract} follows one compiled contract of Adder-4 at $d=13$.
A misprediction skips the $S$ correction of a $T$ gate on $q_0$, which leaves a quarter-turn about $Z_0$.
The walk takes five steps, and four of them cross 12 measurements inside teleported-$S$ and lattice-surgery CNOT gadgets, whose checked identities carry the residual through every outcome.
The first CNOT targets $q_0$, so the quarter-turn spreads to $Z_0Z_8$, with a frame set by the CNOT's outcomes.
The Hadamard gate on $q_8$ turns it into $Z_0X_8$, and the $S$ gate into $Z_0Y_8$.
The recovery script grows with the residual, from one $S$ of 21 rounds on 2 cells to 107 rounds on 10 cells.
The next CNOT on $q_8$ would carry $Z_0Y_8$ out of the library, so the contract ends at a fence.
Contracts that cross measurements are common.
At $d=13$, 56,142 of the programs' 102,582 contracts (54.7\%) cross a measurement.

\section{The Resource-Aware Runtime}\label{sec:runtime}

The recovery contracts give every boundary a recovery script with its length and workspace.
The resource-aware runtime in the controller uses them twice.
First, it reserves the workspace of every script before it speculates on a correction (\textbf{C2}).
Second, it weighs the time that speculating saves against the expected cost of recovery (\textbf{C3}).
We first describe static reservation, then the decision and its tables, and finally the controller's cost and timing.

\subsection{Static reservation}\label{sec:cycle}\label{sec:resv}

\begin{figure}[t]
\centering
\begin{tikzpicture}[x=1mm,y=1mm,font=\sffamily\fontsize{7}{8.4}\selectfont,
  nd/.style={draw=cMute,line width=0.6pt,fill=cMute,rounded corners=0.8mm,align=center,text=white,inner sep=0.7mm,minimum height=5.6mm},
  fq/.style={nd,draw=cBlue,fill=cBlue},
  wf/.style={-{Latex[length=1.6mm,width=1.2mm]},line width=0.7pt,cInk},
  lab/.style={font=\sffamily\fontsize{6.4}{7.6}\selectfont,text=cInk,inner sep=0.4mm,align=center},
  ttl/.style={font=\sffamily\fontsize{7.2}{8.4}\selectfont\bfseries,text=cInk},
  num/.style={circle,fill=cBlue,text=white,font=\sffamily\fontsize{6}{6}\selectfont\bfseries,inner sep=0pt,minimum size=3.2mm}]
  \node[ttl] at (19.5,25.5) {(a) no reservation};
  \node[fq,minimum width=14mm] (ra) at (7.5,18) {$v$'s recovery};
  \node[nd,minimum width=15mm] (ca) at (31.5,18) {patched cell};
  \node[nd] (ia) at (19.5,4) {next instruction};
  \draw[wf] (ra) -- node[lab,above] {needs} (ca);
  \draw[wf] (ca) -- node[lab,right=1.2mm] {freed by} (ia);
  \draw[wf] (ia) -- node[lab,left] {reads $v$} (ra);
  \node[lab,text=cOrange,font=\sffamily\fontsize{6.8}{8}\selectfont\bfseries] at (19.5,12.6) {deadlock};
  \begin{scope}[xshift=44mm]
  \node[ttl] at (19.5,25.5) {(b) static reservation};
  \node[fq,minimum width=14mm] (rb) at (7.5,18) {$v$'s recovery};
  \node[fq,minimum width=15mm] (cb) at (31.5,18) {held cell};
  \node[nd] (ib) at (19.5,4) {next instruction};
  \draw[wf] (rb) -- node[lab,above] {needs} (cb);
  \draw[wf,cMute,dashed] (cb) -- coordinate[pos=0.45] (cut) node[lab,right=1.6mm,pos=0.45,text=cMute] {no patch} (ib);
  \node[text=cOrange,font=\bfseries\fontsize{9}{9}\selectfont,inner sep=0pt] at (cut) {$\times$};
  \draw[wf] (ib) -- node[lab,left=2.6mm,pos=0.55] {waits at\\dispatch} (rb);
  \node[num] at (rb.north west) {1};
  \node[num] at (ib.north west) {2};
  \end{scope}
\end{tikzpicture}
\caption{Without reservation, the recovery of a mispredicted correction $v$ waits for a cell, the cell waits for the next instruction to discard its patch, and that instruction waits for $v$.
  Static reservation holds the cell from registration, so the instruction waits at dispatch and places no patch, which cuts the cycle and lets the recovery run first.}
\label{fig:cycle}
\end{figure}
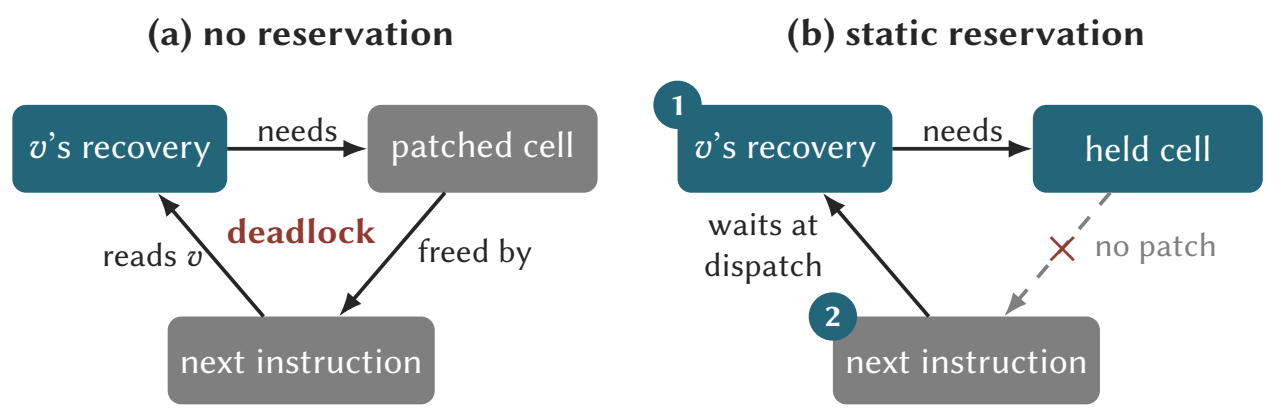

\looseness=-1
Without reservation, one speculative correction suffices to deadlock the program (\secref{sec:problem}).
\figref{fig:cycle}a shows this wait-for cycle.
A pending correction stands at a boundary whose script needs one cell.
An ordinary instruction places a patch on that cell, and only the next instruction on the cell discards the patch.
However, that next instruction reads the qubit that the correction targets, so it waits behind the correction's fence.
After a misprediction, the recovery waits for the cell, the cell waits for the discarding instruction, and that instruction waits for the correction.
One of the two-correction tests below constructs this cycle.

Static reservation removes this cycle, and with it the deadlock of \textbf{C2}, by holding a pending correction's whole footprint, the recovery workspace of every script in its contract, before the correction is speculated on.
The ordinary instruction then waits at dispatch and places no patch, so the recovery runs on the cell, and the instruction runs once the correction settles (\figref{fig:cycle}b).
The runtime manages each pending correction in five operations.
To \textbf{register} a pending correction, the reservation manager reads the footprint as routed on the program's layout and reserves the correction's entries in the write queue, the controller's pending state updates and script writes.
The correction registers only if the write queue has room and no footprint cell is occupied, held or booked.
Otherwise it retries each round until its decode arrives and then waits as SWIPER alone does.
At dispatch, an ordinary instruction that targets a held cell waits until the correction settles.
The runtime \textbf{admits} a step that the decision unit chose only if the step's workspace is free and its light cone overlaps no other live correction's.
The correction \textbf{settles} when its final answer arrives, by dropping its dependency if the answer confirms the prediction and by splicing a recovery script otherwise.
At release, once it has settled, the correction issues its pending writes and returns every held cell.
Because the footprint comes from the contract and the layout alone, every recovery meets the same adjacency and space limits, and static reservation needs no check at run time, no window and no table of checked regions.

\looseness=-1
With its footprint held, nothing that a correction's settlement needs waits on the correction's own fence.
Settlement chooses the shortest usable script at or beyond the end of the correction's started work, and it needs five things.
It needs its workspace clear, its admitted steps complete, its own queue entries, its decode with its script's service, and a usable boundary.
Ordinary work cannot occupy a held cell.
The correction's own steps leave the workspace clear, because a boundary is usable only if, on every cell of its workspace, the last unstarted member that touches the cell removes its patch.
Every outside predecessor of an admitted step has started, and a started instruction completes without waiting.
Service and Delivery (below) supply the decode and the script.
So no wait-for cycle forms, and the correction settles under Service and Delivery alone.

\textbf{Concurrent predictions.}
Several pending corrections may be speculated on at once when their footprints are disjoint.
Registration refuses a cell that another live correction holds, and the runtime refuses a revision whose workspace would leave the footprint.
One dependency still crosses corrections, because another correction's admitted step can place a patch on a held cell.
The runtime therefore admits a step only if none of its unstarted members targets a cell that another live correction holds, and registration treats a cell as busy while another correction's admitted step targets it.
With these two tests, each of the five needs stays inside its correction, so every wait that crosses corrections ends at a correction whose settlement needs nothing outside it.
Sixteen two-correction tests pass on the deployed runtime, and 7 of them fail without the two tests.

\textbf{Service and Delivery.}
\textbf{Service} requires instructions and scripts to end, requests asked every round to be served, and the write queue to have room for the updates of the largest accepted request.
\textbf{Delivery} requires every decode to arrive and every pending update to be written.
The check port arbiter's aging, the script store and the splice engine exist only as estimates (\tabref{tab:cost}), so Service and Delivery stay premises of the design.
Under both premises, every run completes.
For Service, the arbiter serves requests in arrival order, aged requests first, so a request asked every round waits behind at most the 64 requesters of static reservation.
The write queue holds 32,768 writes and drains in 115 rounds, and admission refuses a registration that needs more than 16,384 writes, so an admitted write never waits for room.
A script has at most 64 instructions and ends once the splice engine inserts it.
For Delivery, the reservation manager holds the cores of each reserved decode.

\subsection{Balancing speculation against recovery}\label{sec:decision}\label{sec:prices}

\looseness=-1
With its workspace held, a pending correction can be speculated on, but speculating pays only when the rounds it saves outweigh the expected cost of recovery (\textbf{C3}).
The runtime therefore decides for each pending correction in each round whether to wait, defer or apply.

\looseness=-1
\textbf{State and actions.}
A pending correction's state $(e,b,\beta)$ holds its age $e$, the rounds since its producer ended, the boundary $b$ of its contract at which it stands, and its branch $\beta$, which is $0$ on the uncorrected branch and $1$ once it is applied.
Waiting lasts one round and costs one round, in which the correction's dependents are blocked.
Deferring runs one checked step to the next boundary whose script is placed and whose workspace is free, and its direct cost is the overhead price plus a margin (\secref{sec:select}).
Applying runs a boundary's script early and moves the correction to $(b,1)$.
A correction is applied at most once, only on the uncorrected branch, and only at a boundary whose script is placed and clean and whose contract can be rebased there, that is, re-derived from the applied state.
Applying lasts the script's length $\tau_b$ plus two rounds, from the decision to the script's end.
It costs that time plus its load charge and a price $\xi_b$ for the delay it adds to later decodes, fitted on training runs.

\textbf{Odds.}
Let $\pi_1$ be the probability that the decoded outcome is 1, so that the correction is needed, and $\pi_0=1-\pi_1$.
Branch $\beta$ is right with probability $\pi_\beta$.
A prediction of 1 with confidence $c$ sets $\pi_1=c$, a prediction of 0 sets $\pi_1=1-c$, and a correction with no prediction takes its class's prior rate, $0.505$ when pooled.

\textbf{Recovery price.}
The runtime recovers from a misprediction where execution stands when the decoded outcome arrives.
The price $\rho_\beta(b)$ of this recovery is the script's length at $b$, plus a mean wait before the script starts, plus the rounds until its results are decoded, plus a load charge.
Both waits are static means fitted on training runs, the start wait for each distance, decoder budget and branch, and the decode wait for each recovery class.
The load charge grows with the cores that the script's decodes cannot find free.
When the forecast for the current load bin leaves $\hat F$ cores free and the script's decoding demand is $\nu$ cores, the script meets a shortfall of $s=\max(0,\nu-\hat F)$ cores.
Its load charge is then $\kappa s+\theta U/B$.
Here $B$ is the decoder budget in cores and $\mu$ is the mean full-decode service time of the producer's windows, giving $\kappa=\mu/B$.
The shortfall fraction is $\theta=s/\nu$ when $\nu>0$, and zero otherwise.
The decoding work $U$ sums the mean service times of its decodes.

\looseness=-1
\textbf{Arrival.}
The decoded outcome arrives at an age $e_v$ whose distribution is fitted on the training seed for each decode-latency class, by a Kaplan--Meier estimate that keeps the censored values, with a constant-hazard tail $\bar\Lambda(e)=\Pr[e_v>e]$.

\subsection{The recurrence}\label{sec:recurrence}

The recurrence combines these terms to minimize each correction's expected sum of blocked rounds, recovery rounds, decoding charges and overhead prices.
If the outcome is decoded during action $u$, the recurrence charges the expected recovery at $u$'s destination $(b_u,\beta_u)$.
Otherwise, the correction continues from that destination.
With $g_u$ the direct cost and $d_u$ the duration, the expected cost of $u$ is
\begin{equation}
\begin{aligned}
J_u(e,b,\beta)={}&g_u+\tfrac{\bar\Lambda(e)-\bar\Lambda(e+d_u)}{\bar\Lambda(e)}\,(1-\pi_{\beta_u})\,\rho_{\beta_u}(b_u)\\
&+\tfrac{\bar\Lambda(e+d_u)}{\bar\Lambda(e)}\,J^\ast(e+d_u,b_u,\beta_u),
\end{aligned}
\label{eq:bet}
\end{equation}
and $J^\ast(e,b,\beta)$ is the least $J_u$ over the actions that the contract allows at $(b,\beta)$.
At the fence $F$, only waiting and, on the uncorrected branch, applying remain.
With waiting alone, $J^\ast(e,F,\beta)=M(e)+(1-\pi_\beta)\rho_\beta(F)$, where $M(e)$ is the expected wait for the decode.
The recurrence assumes that a prediction's correctness does not depend on the decode's latency.
Because mispredictions may gather in the slowest decodes, \secref{sec:eval-stress} tests this assumption by placing every misprediction on the slowest decode draws.

\subsection{Settings and the selector}\label{sec:select}

\looseness=-1
Two settings outside the recurrence shape each configuration's decision, a margin on the price of every deferred step and a limit on concurrent predictions.
A selector chooses both on training seeds.

\textbf{The overhead price.}
Even a correct deferred step waits for the scheduler, so each deferred step pays an overhead price.
For each program, distance and budget, one training run with the price off counts the started deferred steps and the rounds a step's own request held its start.
Each deferred step then costs $g=c_o+m$ rounds, with $c_o$ the held rounds per started step and $m$ the margin.

\textbf{The margin and the limit.}
The margin $m$ is $0$, $d$ or $4d$ rounds, or infinite, which waits outright, so a larger margin makes the decision defer less.
The limit allows one prediction at a time, or concurrent predictions up to the hardware's 64 live corrections.
A pending correction that reaches the admission gate at the limit is refused and waits for its decode.
Among each configuration's candidates with the fewest censored runs, waiting wins if its geometric-mean makespan is within $0.5\%$ of the best, and otherwise the smallest geometric mean wins.
With a finite margin, the decision still runs online at every correction and boundary.
Fixed policies, such as always deferring or deferring while the script's volume stays flat, are settings of the same prices, so one selector compares them on equal terms.
When decoder cores are scarcest, the decision itself defers only as far as the second rule allows.
Either rule can also replace only the decision's choice to defer, while the decision's tables still choose between waiting and applying (\secref{sec:eval-lib}).

\subsection{Tables in hardware}\label{sec:tables}

\looseness=-1
To decide every round with one lookup per pending correction, the compiler precomputes \eqref{eq:bet} into tables by a backward sweep over age.
A table belongs to a contract profile, a confidence bin and a load bin.
The profile holds everything the sweep reads, the decode-latency class and the contract up to its cap of 16 boundaries, with step durations, the fence, script lengths and decoding demands, and the rebased profile at each boundary where applying is allowed.
The six confidence bins cover no prediction and five ranges of $\pi_1$, and the three load bins split the forecast of free cores at $5\%$ and $25\%$ of the budget.
For each of eight age bins and 17 boundary lanes, an entry ranks the actions, one of six orders on the uncorrected branch in 3 bits, and defer or wait on the applied branch in 1 bit.

\looseness=-1
Each round, the decision unit makes one lookup per correction, in the row of its profile, confidence bin and age bin at the current load bin.
From the cut, the first boundary whose steps have not begun, it takes at each boundary the first eligible action of the stored order.
It admits one step after another while that action is to defer, and it stops at the first wait or apply.
A later round can take back admitted steps beyond the cut, none of which has begun.
The eligibility checks are hard constraints.
For a deferred step, they cover the fences, pending cone work, disjoint light cones, free workspace, script cells free of other corrections' reservations and the unit's capacity.
An application must be ready, find the decode queue empty or with an expected wait of at most 7 rounds, and find idle cores, net of every pending application, that cover its script's decoding demand.
The checks remove actions but keep the stored costs, which assume that every later action of the contract stays eligible, so the unit's choice of the best-ranked allowed action is a greedy approximation of the best choice under the checks.
The unit holds tables for 64 profiles on each branch, returns an action 8 cycles after a lookup, and serves 64 corrections in at most 32 cycles on its two ports.
In the evaluated programs, the uncorrected branch needed at most 59 profiles, and the applied branch needed more than 64 at 37 of the 50 speculating configurations.
In that case, the compiler keeps the 64 profiles that cover the most pairs of a correction and a boundary where applying is allowed.
The other 1.5\% of these pairs cannot apply, and the simulated runs include this restriction.

\subsection{Cost and timing}\label{sec:hwcost}

\looseness=-1
The decision unit and the reservation manager close timing at a 3.5\,ns clock period on an AMD Versal V80 in out-of-context synthesis.
The rest of the controller is charged at that clock, at which a syndrome round of 1\,\textmu s is 286 cycles.
The decision unit takes 3,146 LUTs.
We derived eleven of the twelve service terms from the block RAM's ports, the banking and the queue's capacity, and estimated the splice engine's.
The read path gives a dependent access of 5 cycles, which every simulated run charges.
A registration takes 18 derived cycles, 6.3\% of a round, and a settlement reaches its script's first instruction in 28 cycles, 9.8\% of a round, of which 18 are in routed blocks and 10 estimated.
In static reservation's 540 uncapped runs at the 45 concurrently speculating configurations (\secref{sec:eval-bounds}), the controller's 32-entry event queue held at most six entries, the 99th percentile of its peak in a round was one, and no controller request waited a round.

Among the estimated blocks, the script store's 64 URAM288 blocks hold every script, each correction's script table and the tables rebased for applying at each boundary up to a limit per program, past which applying fetches the table from HBM.
No run of the main comparison (\secref{sec:eval-main}) fetched a table from HBM.
The hardware design's sizing method estimates 21 to 60 blocks by program.

\begin{table}[t]
\caption{Controller's cost on the V80 sums the routed part, the estimated part and the part counted by unit cost.}
\label{tab:cost}
\centering
\setlength{\tabcolsep}{4pt}
\resizebox{\columnwidth}{!}{%
\begin{tabular}{@{}lll@{}}
\toprule
\textbf{Block} & \textbf{Memory blocks} & \textbf{LUTs (k)} \\
\midrule
static reservation & \textbf{21} RAMB36 & \textbf{18.7} + 30.5 (22.8--{}54.2) + \textit{3.1} \\
script store & 21 to 60 URAM288 & \textit{0.5} \\
splice engine & 2 (1.5--{}2) RAMB36 & 1.0 (1.0--{}2.0) + \textit{1.8} \\
\bottomrule
\end{tabular}}
\end{table}

In \tabref{tab:cost}, static reservation's 21 RAMB36 blocks are all in the routed decision unit, and its largest estimate is the cell state.
Waiting needs none of this hardware, so its cost adds to the qubits and decoder cores that every comparison with waiting holds equal.

\section{Methodology}\label{sec:method}

We evaluate \sys{} in simulation against SWIPER alone, the strongest baseline we have, because SWIPER's speculation decodes several layers of dependent windows in parallel.
Because a faster decoder leaves shorter waits, this comparison leaves \sys{} the least waiting to remove.
We experiment on eight programs at 88 configurations, and every simulated run includes the controller's service times.

\textbf{Simulator.}
Each system runs in SWIPER-SIM~\cite{swiper} through an adapter, in the simulator's hardware-limit mode.
Both systems share the scheduler, the random streams, the decoder timing, the layout and on-demand magic states, and the scheduler includes the latency of the control path.
Every program runs on the EDPC layout~\cite{beverland2022edpc} of SWIPER's compilation chain~\cite{leblond2024}, where lattice surgery joins two tiles only along a routing path between them.
The same router places each recovery script on the program's layout, so the cells that a script reserves include its routing paths.
A contract therefore ends at a fence before a step whose script the router cannot place.

\looseness=-1
\textbf{Modeled inputs.}
The simulator draws the decode time of each window from a latency table that we measured with SWIPER's script~\cite{swiper}.
We built this table from PyMatching~2.4.0~\cite{pymatching} decodes of stim's~\cite{stim} rotated memory circuits under circuit-level noise of $10^{-3}$, at distances 3 to 21 and windows of 1 to 7 times $d$ rounds, with 10,000 shots each.
We timed every decode alone on one core of an idle AMD Ryzen~9 9950X.
The predictor's accuracy in simulation is the lower bound of its measured accuracy (\secref{sec:client}), drawn independently of the decode times.
The decision unit and the reservation manager use their synthesized timing, and the other controller blocks use the service times that \secref{sec:hwcost} derives.
In both systems, each round of a Hadamard gate takes two ordinary rounds, which is the time that its sampled circuit takes.

\begin{table}[t]
\caption{Programs and peak decoder demand in cores.}
\label{tab:programs}
\centering
\setlength{\tabcolsep}{4pt}
\resizebox{\columnwidth}{!}{%
\begin{tabular}{@{}lrrrlrrr@{}}
\toprule
\textbf{Program} & \textbf{Qubits} & \textbf{{\boldmath$T$} gates} & \textbf{{\boldmath$T$} depth} & \textbf{Feedback} & \textbf{{\boldmath$d=13$}} & \textbf{17} & \textbf{21} \\
\midrule
QPE-5 & 5 & 1,813 & 957 & $S$ & 22 & 30 & 43 \\
QPE-10 & 10 & 8,644 & 2,520 & $S$ & 47 & 61 & 84 \\
QFT-10 & 10 & 9,428 & 2,466 & $S$ & 46 & 64 & 84 \\
Grover-5 & 5 & 14,085 & 9,699 & $S$ & 25 & 35 & 45 \\
Adder-4 & 11 & 12 & 4 & $S$, $CZ$ & 39 & 57 & 86 \\
Adder-8 & 23 & 28 & 8 & $S$, $CZ$ & 57 & 88 & 143 \\
Adder-16 & 47 & 60 & 16 & $S$, $CZ$ & 114 & 172 & 277 \\
Adder-32 & 95 & 124 & 32 & $S$, $CZ$ & 259 & 362 & 506 \\
\bottomrule
\end{tabular}}
\end{table}

\textbf{Programs and configurations.}
Four programs come from MQT Bench~\cite{mqtbench}, \ie, quantum phase estimation on 5 and 10 qubits (QPE-5, QPE-10), the quantum Fourier transform on 10 (QFT-10) and Grover search on 5 (Grover-5), compiled by the Lattice Surgery Compiler~\cite{lsqecc}.
The other four are Gidney's temporary-AND adders~\cite{gidney2018} on 4, 8, 16 and 32 bits.
In every program, feedback conditions the $T$ gadget's $S$, and in the adders also the uncomputation's $CZ$ (\tabref{tab:programs}).
As in SWIPER~\cite{swiper}, we compile each $T$ gate with a conditional $S$ correction, so both systems run on the same layout and patches.
Neither system uses auto-corrected rotations~\cite{litinski}.
A configuration is one program at one code distance and decoder budget.
At a level $\ell$ of $50$, $75$, $100$ or $150\%$, the decoder budget is $\lceil \ell P\rceil$ cores, where the peak decoder demand $P$ is the largest number of decoders in use at once in one run of SWIPER alone with unlimited decoders.

\looseness=-1
As \tabref{tab:programs} shows, $P$ ranges from 22 cores for QPE-5 at $d=13$ to 506 for Adder-32 at $d=21$.
The 88 configurations are the eight programs at $d=13$ and $17$ at all four budgets, and at $d=21$ at all but $50\%$, where earlier runs of both systems were limited by the decoder and exceeded the censoring bound below.

\looseness=-1
\textbf{Systems.}
Both systems decode with SWIPER's speculative window decoding at its default settings with a speculation accuracy of 0.9 and one round of speculation time.
SWIPER alone runs this decoder without \sys{}, so no operation proceeds on a predicted correction.
\sys{} runs on top of the same decoder with the recovery library, static reservation, the local-decoding predictor and at most 64 live corrections.
At each configuration, \sys{} uses the policy settings that the training-time selector of \secref{sec:select} chose on ten training seeds, with prices fitted on the same training runs.
The selected settings allow concurrent predictions at 45 configurations, one prediction at a time at 5, and waiting at 38.
Of these 50 speculating configurations, 33 run the decision at a margin of 0, 10 at $d$ rounds and 7 at $4d$ rounds.
At a waiting configuration, the two systems behave identically by construction, so we do not run it.

\textbf{Metric and statistics.}
We measure quantum-side time as the makespan, the number of rounds until every pending correction has settled.
For each of the 50 speculating configurations, each system ran on twelve test seeds that no training run used.
We censor a run whose makespan exceeds four times that of SWIPER alone, and all 1,200 runs of the two systems completed within this bound.

For each seed, our statistic over a set of configurations is the mean log ratio of the two systems' makespans.
A comparison's ratio $r$ is the exponential of this statistic's mean over the twelve seeds, with a two-sided $95\%$ interval.
Each primary comparison is a one-sided $t$ test with 11 degrees of freedom against a ratio of 1, and the two primary comparisons form one family under Holm's procedure at a family-wise $0.05$.
The training-time selection fixed every configuration set and never read a test seed.
The 38 waiting configurations contribute a log ratio of 0, so over the 88 configurations each seed's statistic is $50/88$ of its value over the 50, and the comparisons over the 88 share the $t$ and $p$ of the 50.

\section{Evaluation}\label{sec:eval}

We evaluate \sys{} to answer four questions.
\ding{182} How much quantum-side time does \sys{} reduce against SWIPER alone? (\secref{sec:eval-main})
\ding{183} What explains the reduction, and does the runtime work as designed? (\secref{sec:eval-lib})
\ding{184} Is \sys{} safe, and what does its safety cost? (\secref{sec:eval-bounds})
\ding{185} How sensitive is the reduction to the controller, the decoder and the predictions? (\secref{sec:eval-stress})

\begin{figure}[t]
\centering
\resizebox{\columnwidth}{!}{%
\begin{tikzpicture}
\begin{axis}[width=\columnwidth, height=3.6cm, ybar, bar width=5.4pt,
  xmin=0.45, xmax=8.55, ymin=-2.5, ymax=31, xtick={1,...,8},
  xticklabels={Grover-5,QFT-10,QPE-10,QPE-5,Adder-4,Adder-8,Adder-16,Adder-32},
  xticklabel style={font=\scriptsize, rotate=30, anchor=north east, inner sep=1pt},
  major x tick style={draw=none}, ytick={0,10,20,30}, yticklabel style={font=\footnotesize},
  ylabel={quantum-side time reduction\\against SWIPER alone (\%)}, ylabel style={font=\footnotesize, align=center, yshift=-4pt},
  axis x line*=bottom, axis y line*=left, axis line style={cMute}, tick style={cMute}, ymajorgrids,
  grid style={cGrid,line width=0.3pt},
  legend style={at={(0.99,0.98)}, anchor=north east, draw=none, fill=none, font=\scriptsize, legend columns=1,
    row sep=-1pt},
  legend image code/.code={\fill[#1] (0pt,-2pt) rectangle (7pt,3.5pt);}]
  \addplot[fill=cBlue!50, draw=none, bar shift=-6.4pt, error bars/.cd, y dir=both, y explicit, error mark=none,
    error bar style={cInk,line width=0.5pt}]
    table[x=x, y=c13, y error plus expr=\thisrow{hi13}-\thisrow{c13}, y error minus expr=\thisrow{c13}-\thisrow{lo13}] {data/cutbars.dat};
  \addplot[fill=cBlue!75, draw=none, bar shift=0pt, error bars/.cd, y dir=both, y explicit, error mark=none,
    error bar style={cInk,line width=0.5pt}]
    table[x=x, y=c17, y error plus expr=\thisrow{hi17}-\thisrow{c17}, y error minus expr=\thisrow{c17}-\thisrow{lo17}] {data/cutbars.dat};
  \addplot[fill=cBlue, draw=none, bar shift=6.4pt, error bars/.cd, y dir=both, y explicit, error mark=none,
    error bar style={cInk,line width=0.5pt}]
    table[x=x, y=c21, y error plus expr=\thisrow{hi21}-\thisrow{c21}, y error minus expr=\thisrow{c21}-\thisrow{lo21}] {data/cutbars.dat};
  \legend{$d=13$, $d=17$, $d=21$}
  \draw[cInk,line width=0.6pt] (axis cs:0.45,0) -- (axis cs:8.55,0);
  \foreach \s in {-6.4pt, 6.4pt} {
    \edef\tmp{\noexpand\node[rotate=90, anchor=west, font=\noexpand\scriptsize, text=cMute, inner sep=0.5pt]
      at ([xshift=\s]axis cs:5,0.4) {waits};}\tmp}
  \node[rotate=90, anchor=west, font=\scriptsize, text=cMute, inner sep=0.5pt] at ([xshift=6.4pt]axis cs:6,0.4) {waits};
  \foreach \p in {7,8} {
    \edef\tmp{\noexpand\node[anchor=south, font=\noexpand\scriptsize, text=cMute, inner sep=0.5pt]
      at (axis cs:\p,0.4) {waits};}\tmp}
\end{axis}
\end{tikzpicture}}
\caption{Reduction of quantum-side time per program and code distance with $95\%$ intervals. ``waits'' marks a program and distance where \sys{} waits in every configuration.}
\label{fig:cuts}
\end{figure}
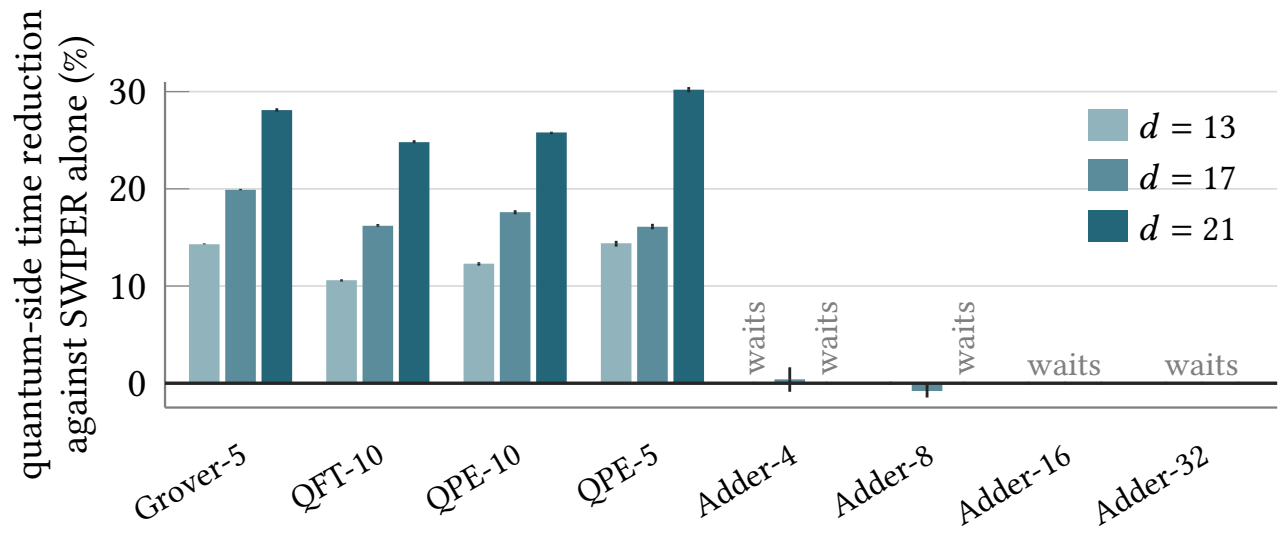

\subsection{Reduction of quantum-side time}\label{sec:eval-main}\label{sec:eval-where}

\looseness=-1
\sys{} reduced quantum-side time against SWIPER alone by 9.81\% over all 88 configurations.
This average includes the 38 configurations where \sys{} waits, among them every configuration of Adder-16 and Adder-32, which run alike in both systems and show no reduction by construction.
Over the 50 speculating configurations, the reduction was 16.62\%, and over the 45 concurrently speculating ones it reached 18.34\%.
The reduction peaked at 31.1\% for QPE-5 at $d=21$, at budgets of 100 and 150\% of peak demand.

The comparisons under all settings are significant under Holm's procedure, each with $p<10^{-23}$ in its one-sided test.
These results show that speculating on live quantum states with in-place recovery shortens execution even against the strongest decoder baseline we have.
The training-time selector confines speculation to the configurations where it is expected to pay, so the other 38 run exactly as SWIPER alone does, and the gain is largest where \sys{} speculates on several pending corrections at once.

\figref{fig:cuts} breaks down the reduction by program and code distance.
Grover-5, QFT-10, QPE-10 and QPE-5 account for the reduction, with 10.6 to 30.2\% each.
For each of these four programs, the reduction increases with code distance, from 10.6--14.4\% at $d=13$ to 24.8--30.2\% at $d=21$.
The 30.2\% of QPE-5 at $d=21$ is the geometric mean over three decoder budgets, the largest of which yields the peak reduction of 31.1\%.
In contrast, the adders show little or no reduction.
Adder-4 speculates only at $d=17$, where its 0.4\% reduction is not significant, because its $95\%$ interval includes zero.
The selector chose to speculate on Adder-8 at $d=13$ and $17$, one prediction at a time, because waiting was 0.52--0.73\% slower on the training seeds.
On the test seeds, however, Adder-8 showed no significant change at $d=13$ and a 0.8\% slowdown at $d=17$.

\looseness=-1
We attribute this to the adders' few $T$ gates, 12 to 124 against 1,813 to 14,085 in the other programs (\tabref{tab:programs}), which leave little to speculate on, and to their measurement-conditioned $CZ$, which always waits for its decode (\secref{sec:client}).
Overall, \sys{} benefits programs with many $T$ gates the most, and its benefit grows with code distance, because longer decodes leave more waiting to remove.

\subsection{Understanding the reduction}\label{sec:eval-lib}
To explain the reduction, we examine how the decision speculated, how much waiting \sys{} removed, how the decision compares with fixed rules, whether the runtime's cost estimates were accurate, and whether the reduction depends on extra workspace.

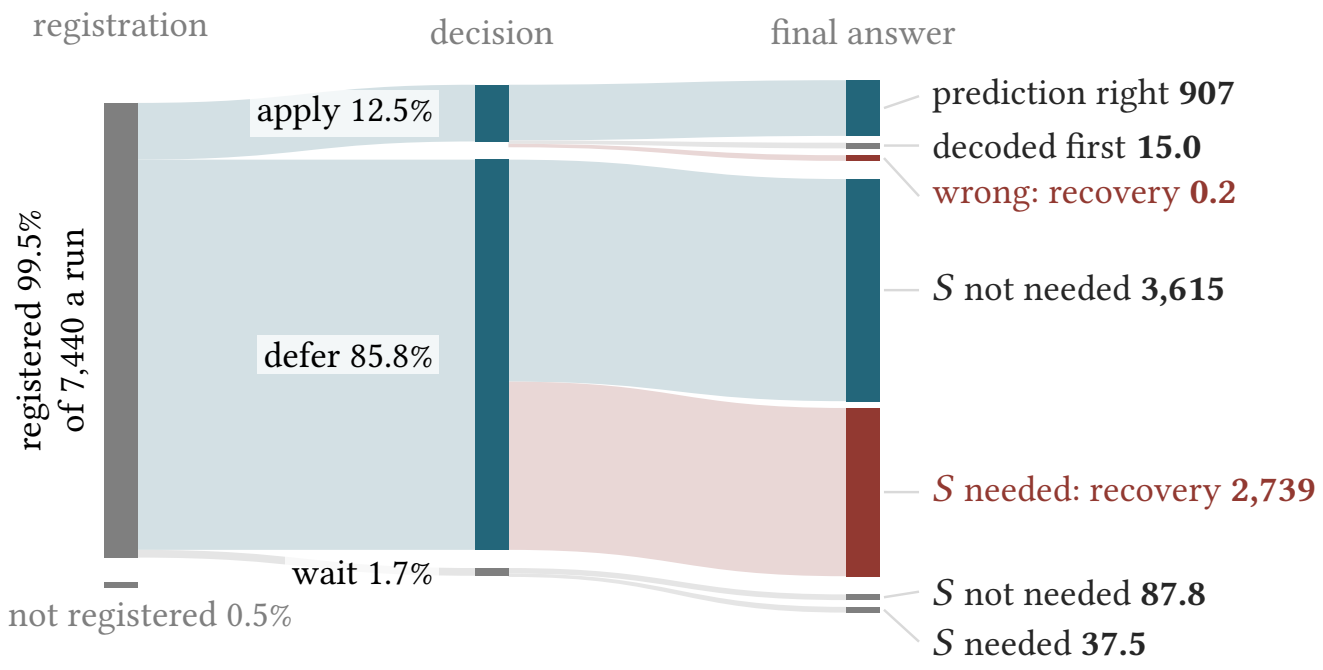
\begin{figure}[t]
\centering
\resizebox{\columnwidth}{!}{%
\begin{tikzpicture}[x=1.18cm,y=0.65cm,font=\scriptsize,every node/.style={inner sep=1pt}]
\useasboundingbox (0.04,-0.95) rectangle (6.25,4.85);
\fill[cBlue!20] (0.710,4.000) .. controls (1.530,4.000) and (1.530,4.160) .. (2.350,4.160) -- (2.350,3.661) .. controls (1.530,3.661) and (1.530,3.501) .. (0.710,3.501) -- cycle;
\fill[cBlue!20] (0.710,3.501) .. controls (1.530,3.501) and (1.530,3.501) .. (2.350,3.501) -- (2.350,0.068) .. controls (1.530,0.068) and (1.530,0.068) .. (0.710,0.068) -- cycle;
\fill[cMute!20] (0.710,0.068) .. controls (1.530,0.068) and (1.530,-0.092) .. (2.350,-0.092) -- (2.350,-0.160) .. controls (1.530,-0.160) and (1.530,-0.000) .. (0.710,-0.000) -- cycle;
\fill[cBlue!20] (2.510,4.160) .. controls (3.330,4.160) and (3.330,4.200) .. (4.150,4.200) -- (4.150,3.710) .. controls (3.330,3.710) and (3.330,3.670) .. (2.510,3.670) -- cycle;
\fill[cMute!20] (2.510,3.670) .. controls (3.330,3.670) and (3.330,3.650) .. (4.150,3.650) -- (4.150,3.600) .. controls (3.330,3.600) and (3.330,3.640) .. (2.510,3.640) -- cycle;
\fill[cOrange!20] (2.510,3.640) .. controls (3.330,3.640) and (3.330,3.540) .. (4.150,3.540) -- (4.150,3.490) .. controls (3.330,3.490) and (3.330,3.610) .. (2.510,3.610) -- cycle;
\fill[cBlue!20] (2.510,3.501) .. controls (3.330,3.501) and (3.330,3.330) .. (4.150,3.330) -- (4.150,1.376) .. controls (3.330,1.376) and (3.330,1.548) .. (2.510,1.548) -- cycle;
\fill[cOrange!20] (2.510,1.548) .. controls (3.330,1.548) and (3.330,1.316) .. (4.150,1.316) -- (4.150,-0.164) .. controls (3.330,-0.164) and (3.330,0.068) .. (2.510,0.068) -- cycle;
\fill[cMute!20] (2.510,-0.092) .. controls (3.330,-0.092) and (3.330,-0.324) .. (4.150,-0.324) -- (4.150,-0.374) .. controls (3.330,-0.374) and (3.330,-0.140) .. (2.510,-0.140) -- cycle;
\fill[cMute!20] (2.510,-0.140) .. controls (3.330,-0.140) and (3.330,-0.434) .. (4.150,-0.434) -- (4.150,-0.484) .. controls (3.330,-0.484) and (3.330,-0.170) .. (2.510,-0.170) -- cycle;
\fill[cMute] (0.550,4.000) rectangle (0.710,0.000);
\fill[cMute] (0.550,-0.220) rectangle (0.710,-0.270);
\fill[cBlue] (2.350,4.160) rectangle (2.510,3.661);
\fill[cBlue] (2.350,3.501) rectangle (2.510,0.068);
\fill[cMute] (2.350,-0.092) rectangle (2.510,-0.160);
\fill[cBlue] (4.150,4.200) rectangle (4.310,3.710);
\fill[cMute] (4.150,3.650) rectangle (4.310,3.600);
\fill[cOrange] (4.150,3.540) rectangle (4.310,3.490);
\fill[cBlue] (4.150,3.330) rectangle (4.310,1.376);
\fill[cOrange] (4.150,1.316) rectangle (4.310,-0.164);
\fill[cMute] (4.150,-0.324) rectangle (4.310,-0.374);
\fill[cMute] (4.150,-0.434) rectangle (4.310,-0.484);
\node[anchor=south,text=cMute] at (0.630,4.450) {registration};
\node[anchor=south,text=cMute] at (2.430,4.450) {decision};
\node[anchor=south,text=cMute] at (4.230,4.450) {final answer};
\node[anchor=south,rotate=90,align=center] at (0.510,2.000) {registered 99.5\%\\of 7,440 a run};
\node[anchor=north west,align=left,text=cMute] at (0.050,-0.310) {not registered 0.5\%};
\node[anchor=east,fill=white,fill opacity=0.75,text opacity=1] at (2.180,3.911) {apply 12.5\%};
\node[anchor=east,fill=white,fill opacity=0.75,text opacity=1] at (2.180,1.785) {defer 85.8\%};
\node[anchor=east,fill=white,fill opacity=0.75,text opacity=1] at (2.180,-0.126) {wait 1.7\%};
\draw[cGrid] (4.330,3.955) -- (4.500,4.060);
\node[anchor=west,text=cInk] at (4.530,4.060) {prediction right \textbf{907}};
\draw[cGrid] (4.330,3.625) -- (4.500,3.625);
\node[anchor=west,text=cInk] at (4.530,3.625) {decoded first \textbf{15.0}};
\draw[cGrid] (4.330,3.515) -- (4.500,3.180);
\node[anchor=west,text=cOrange] at (4.530,3.180) {wrong: recovery \textbf{0.2}};
\draw[cGrid] (4.330,2.353) -- (4.500,2.353);
\node[anchor=west,text=cInk] at (4.530,2.353) {$S$ not needed \textbf{3,615}};
\draw[cGrid] (4.330,0.576) -- (4.500,0.576);
\node[anchor=west,text=cOrange] at (4.530,0.576) {$S$ needed: recovery \textbf{2,739}};
\draw[cGrid] (4.330,-0.349) -- (4.500,-0.300);
\node[anchor=west,text=cInk] at (4.530,-0.300) {$S$ not needed \textbf{87.8}};
\draw[cGrid] (4.330,-0.459) -- (4.500,-0.740);
\node[anchor=west,text=cInk] at (4.530,-0.740) {$S$ needed \textbf{37.5}};
\end{tikzpicture}}
\caption{Outcomes of the pending corrections in a \sys{} run.}
\label{fig:flow}
\end{figure}

\figref{fig:flow} shows where the pending corrections of a \sys{} run end up, averaged over the 50 speculating configurations.
Of the 7,440 pending corrections in a run, 99.5\% were registered, and the decision speculated on almost all of them, deferring 85.8\% and applying 12.5\% while waiting on only 1.7\%.
Of the 0.5\% that were not registered, 72.0\% were decoded before they could be registered, 25.6\% were rejected because work in their light cones had already started when the prediction arrived, and 2.3\% were rejected by the limit of one prediction at a time, while the capacity of 64 live corrections rejected none.
Early applications were almost always right, with 0.2 mispredictions per run out of about 922.
Of the 6,354 deferred corrections per run, 2,739 (43\%) later needed $S$ and ran a recovery, so most recoveries follow deferrals, and mispredicted applications cause almost none.
Deferring alone still reduces quantum-side time by 7.0\% when no prediction is available (\secref{sec:eval-stress}).

\sys{} removed 61\% of the waiting for pending corrections.
For every round, the simulator records each instruction whose start is delayed and the reason for the delay.
Over the 50 speculating configurations, the instruction-rounds delayed by pending corrections under \sys{} were 39\% of those under SWIPER alone.
Of these remaining instruction-rounds, 56.5\% were delayed at the end of a contract, beyond which \sys{} cannot speculate, 31.9\% waited for a recovery script, 0.3\% waited for corrections that were not registered, and the other 11.3\% waited where the decision chose to wait before the contract's end (6.1\%), for the admission of a step that the decision allowed (4.8\%), or for a cell that a recovery script held or a measurement-conditioned $CZ$ (0.3\%).
The largest share of the remaining delay thus comes from the length of contracts, which suggests that longer contracts would reduce it further.

Prior work~\cite{spequlation} chooses how far to run ahead with fixed strategies.
To compare with such strategies, at the 39 of the 50 speculating configurations where the selector chose the decision, we let a fixed rule choose when to defer and kept the decision's tables for waiting and applying.
With one prediction at a time and three seeds, the decision reduced quantum-side time by 11.1\%.
Always deferring reduced it by only 10.0\%, and paired with the decision's runs it took 1.25\% longer (interval 0.64 to 1.87\%).
Deferring while the script's volume stays flat from one boundary to the next, a rule that reads the same recovery contracts as the decision, reduced it by 10.9\%, and its paired difference of 0.22\% (interval $-0.03$ to 0.46\%) is not significant.
The decision thus outperforms a rule that ignores the cost of recovery and matches one that reads the same contracts.

The recovery price that the decision uses closely matched the measured recovery time, so the decision weighs speculation against an accurate estimate (\textbf{C3}).
For the recoveries on the uncorrected branch, we compared the price with the measured rounds from the final answer until the recovery's results were decoded.
Their ratio was 1.07, the geometric mean over the 48 configurations with such recoveries, with per-configuration ratios from 0.87 to 1.47.
The price underestimated the measured time by 8.8\% at $d=13$ and overestimated it by 16.0\% at $d=17$ and 13.9\% at $d=21$.

\looseness=-1
The reduction does not come from the extra workspace that static reservation holds.
To test this, we kept the layout unchanged and capped the occupied and held cells of both systems alike at $d=13$, either at the median peak of SWIPER alone or midway between that and \sys{}'s peak.
Under the two caps, the settings for one prediction at a time still reduced quantum-side time by 7.88\% and 7.90\%, and the settings that allow concurrent predictions by 12.85\% and 12.86\%.

\subsection{Safety and its cost}\label{sec:eval-bounds}

\sys{} guarantees that every recovery script is checked and that no recovery deadlocks.
This subsection examines what these guarantees cost in workspace, in run time and in compile time.

\textbf{Workspace cost.} Static reservation holds the cells of every script, so \sys{} holds more workspace than waiting does.
Over the 50 speculating configurations, the most cells in use at once, occupied by the program or held for a recovery, averaged 21.5 under \sys{} and 19.9 under SWIPER alone, so \sys{} needed 1.6 more cells on average and at most 4.3 more, for Grover-5 at $d=21$.

\textbf{Run time cost.} Static reservation may delay execution, because it holds a correction's whole footprint from registration, including cells that the correction never uses.
To measure this delay, we compared it with a simpler admission that holds only the workspace of the boundary where the correction stands.
The simpler admission lets a correction advance only onto workspace that no other correction holds and no admitted instruction has booked, but it does not rule out the wait-for cycle of \secref{sec:cycle}.
A runtime that rules out this cycle at every step needs about 13 times the logic of static reservation, 241.1\,k against 18.7\,k LUTs.
With concurrent predictions, the simpler admission reduced quantum-side time by 1.19\% over the 45 concurrently speculating configurations (interval 1.11 to 1.27\%), and with one prediction at a time, it reduced quantum-side time by 0.10\% over all 88 (interval 0.06 to 0.15\%).
The run-time cost of the no-deadlock guarantee is therefore only 1.19\% with concurrent predictions and 0.10\% with one prediction at a time.

\textbf{Compile time cost.} The compiler checks every recovery script by the checks in \secref{sec:cert}.
All 77,295 scripts that our evaluation used passed this check.
Generating the scripts and profiles took at most 471\,s per program and distance on a single core.

\subsection{Sensitivity}\label{sec:eval-stress}\label{sec:eval-engine}

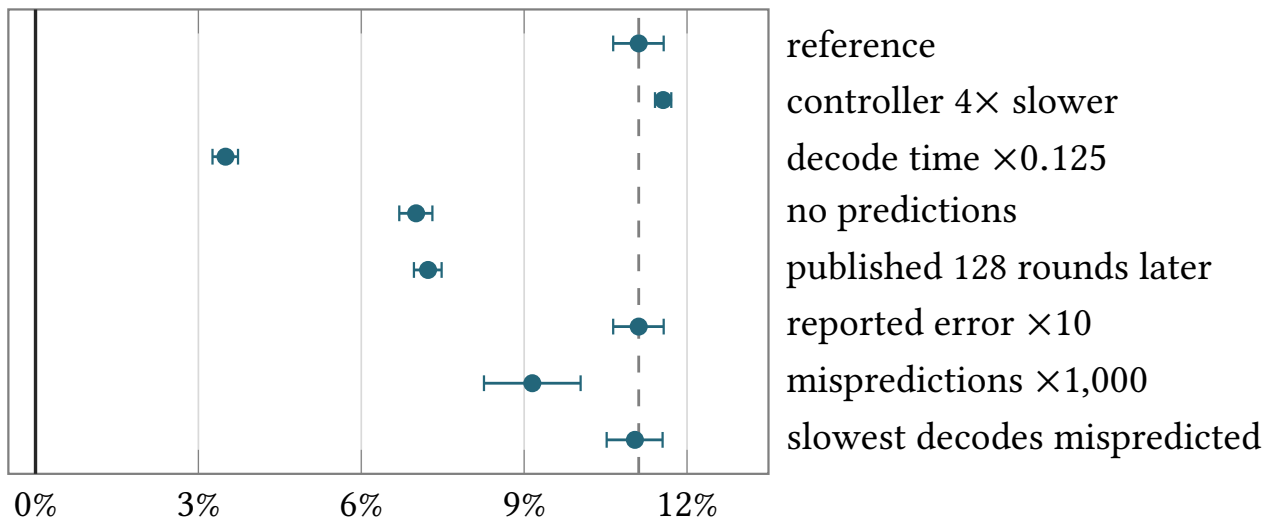
\begin{figure}[t]
\centering
\resizebox{\columnwidth}{!}{%
\begin{tikzpicture}
\begin{axis}[width=0.56\columnwidth, height=2.9cm, scale only axis,
  yticklabel pos=right, yticklabel style={font=\footnotesize, align=left},
  xmin=-0.5, xmax=13.5, ymin=0.4, ymax=8.6,
  xtick={0,3,6,9,12}, xticklabels={0\%,3\%,6\%,9\%,12\%},
  xticklabel style={font=\footnotesize},
  ytick={1,...,8},
  yticklabels={slowest decodes mispredicted, mispredictions $\times 1{,}000$, reported error $\times 10$,
    published 128 rounds later, no predictions, decode time $\times 0.125$, controller 4$\times$ slower,
    reference},
  major y tick style={draw=none},
  axis line style={cMute}, tick style={cMute}, xmajorgrids, grid style={cGrid,line width=0.3pt}]
  \draw[cInk,line width=0.6pt] (axis cs:0,0.4) -- (axis cs:0,8.6);
  \draw[cMute,dashed,line width=0.5pt] (axis cs:11.11,0.4) -- (axis cs:11.11,8.6);
  \addplot+[only marks, mark=*, mark size=1.4pt, color=cBlue, mark options={fill=cBlue},
    error bars/.cd, x dir=both, x explicit, error bar style={line width=0.5pt}]
    table[row sep=\\, x=r, y=y, x error plus expr=\thisrow{hi}-\thisrow{r}, x error minus expr=\thisrow{r}-\thisrow{lo}] {
    y r lo hi \\
    8 11.11 10.64 11.57 \\
    7 11.56 11.41 11.71 \\
    6 3.50 3.26 3.73 \\
    5 7.01 6.70 7.31 \\
    4 7.23 6.97 7.48 \\
    3 11.11 10.64 11.57 \\
    2 9.15 8.26 10.04 \\
    1 11.04 10.52 11.55 \\
  };
\end{axis}
\end{tikzpicture}}
\caption{Sensitivity of \sys{}'s reduction of quantum-side time against SWIPER alone at the most extreme setting of each stress, with $95\%$ intervals. Higher is better. The dashed line shows the reference.}
\label{fig:stress}
\end{figure}

We measured how \sys{}'s reduction responds to changes in the controller, the decoder and the predictions.
\figref{fig:stress} shows the most extreme setting of each type of change.
Note that the reference reduction of 11.1\% is lower than the 16.62\% of \secref{sec:eval-main} because these runs use the settings chosen for one prediction at a time.

\textbf{Controller and decoder.}
\sys{} tolerates a slower controller, and its reduction shrinks but persists with a faster decoder.
With every controller service time two times slower, the reduction was 10.9\%.
At four times slower, one of Adder-8's controller operations no longer fits its port in a round, so Adder-8 cannot run, and over the other 45 configurations the reduction was 11.6\%, against 12.4\% at the reference.
When every decode, including the predictor's, took a half, a quarter or an eighth of its time in both systems, the reduction fell to 7.3\%, 4.9\% and 3.5\%, respectively.
The reduction shrinks because a faster decoder leaves less waiting to remove, yet it remains positive even with decodes eight times faster.

\textbf{Predictions.}
\sys{} degrades gracefully when predictions are fewer, later, less accurate or from a weaker source.
With predictions for half of the pending corrections, the reduction was 9.1\%, and with no predictions it was still 7.0\%, because \sys{} still deferred steps ahead of their decodes.
A predicted correction can be applied early only between the prediction's arrival and its decode, so later predictions gain less.
Publishing every prediction 16, 64 and 128 rounds later gave reductions of 10.6\%, 8.6\% and 7.2\%.

\looseness=-1
SWIPER's unconfirmed window decode can also serve as a prediction source, but its prediction arrives only 2.5 rounds before SWIPER confirms it, against 32.9 rounds for the local-decoding predictor.
With this source, \sys{} ran 3.92\% longer than with the local-decoding predictor over the 50 speculating configurations (interval 3.80 to 4.04\%), yet it still reduced quantum-side time against SWIPER alone by 4.43\% over all 88.

\looseness=-1
\sys{} is also robust to miscalibrated confidence and to frequent mispredictions.
When the predictor reported error probabilities ten times larger or smaller than its actual ones, the reduction stayed at 11.1\%.
When mispredictions became 10 or 100 times as frequent, the reduction was 11.1\% and 10.8\%, whether the reported confidence stayed the same or was corrected to the new rate.
Even when mispredictions were 1,000 times as frequent, a misprediction rate of about 31\%, the reduction was still 9.2\% with the original confidence and 9.1\% with the corrected one.

\section{Discussion}\label{sec:discussion}

\textbf{Scope of the evidence.}
Our comparisons are simulated, with the controller's service times derived from its routed and estimated blocks, and both systems run at the same code distance.
The compiler's check shows that each recovery is exact under ideal instruments, but we did not measure the logical failures that recovery adds under noise.
A first-order estimate suggests that \sys{} reduces this exposure overall.
It assumes that, at a fixed code distance, logical failures grow in proportion to patch-rounds, the syndrome rounds of all occupied patches, idle or active.
Over the 50 speculating configurations, the recovery scripts added 229,483 patch-rounds per run, an upper bound because a script counts whole even on an otherwise idle patch.
Because \sys{} also shortened execution, the other patch-rounds fell by 1,025,472 per run, so the total was 9\% lower than under SWIPER alone.
The total was higher only in 29 of the 600 runs, all of Adder-4 and Adder-8, and by at most 0.9\%, at Adder-8's three configurations at $d=17$.
A comparison at a common logical failure target still needs a failure budget for the recovery scripts, which we will build by charging each script's gates at measured error rates.

\textbf{Generality.}
\sys{} does not depend on a particular decoder or predictor, because any decoder that confirms a decoded outcome and any source that supplies the three items of \secref{sec:client} fit its interface.
It applies to any pending correction whose misprediction leaves residuals in the recovery library, which here holds single quarter-turns on at most two qubits.
The conditional $S$ of a $T$ gadget meets this condition, and so would other teleportation gadgets whose conditional correction is a Clifford gate of the same form.
Corrections with larger residuals, such as the measurement-conditioned $CZ$ of the adders, need a library that holds three quarter-turns and a more accurate predictor, which we leave to future work.
The compiler derives every recovery contract from gadget templates, so another code or layout would need templates for its own gadgets.

\textbf{Scale and hardware.}
Our programs use at most 95 logical qubits and code distances up to 21.
Larger programs may lengthen contracts and need more profiles than the decision unit's 64 per branch, which already leave 1.5\% of the correction-boundary pairs where applying is allowed unable to apply (\secref{sec:tables}).
We will evaluate \sys{} on larger programs and code distances.
In addition, only the decision unit and the reservation manager are synthesized, and the service times of the other controller blocks are derived or estimated (\secref{sec:hwcost}).
We will implement the full controller on an FPGA alongside a real-time decoder and measure its timing in the loop.

\section{Related Work}\label{sec:related}

The closest works speculate inside the classical decoder, run quantum operations ahead, recover from mispredictions in processors, allocate lattice-surgery resources online, or shorten the decode.

\textbf{Speculation inside the decoder.}
Sliding-window decoding decodes a window once the windows it depends on are decoded~\cite{skoric2023,tan2023}.
SWIPER guesses window boundaries so that dependent windows decode in parallel~\cite{swiper}.
Its gain depends on gate speed, speculation accuracy and decoder latency~\cite{li2026specwin}.
Similarly, Malarchick's compiler pass sets which windows to restart after a misprediction~\cite{malarchick2026verifier}.
However, these schemes never run the program ahead.
\sys{} works above them, and we run it on top of SWIPER.

\textbf{Running quantum operations ahead.}
Clifford-frame deferral holds a skipped correction in a frame that later operations read~\cite{chamberland2018}, and \sys{} includes that option inside its decision.
ARTERY predicts a readout that conditions feedback and pre-executes the predicted branch~\cite{artery}.
Similarly, pipelined execution lets successor operations start before their predecessors finish decoding, rolls part of the work back when a prediction fails, and compares fixed strategies for how far to run ahead~\cite{spequlation}.
\sys{} recovers from a misprediction with a recovery script checked for the boundary that execution has reached, holds the script's workspace in advance, and decides for each pending correction whether speculating pays.

\looseness=1
\textbf{Speculation in processors.}
A processor recovers from a mispredicted branch by discarding speculative state, which a reorder buffer keeps apart from the architectural state~\cite{smith1988}, and branch predictors supply the guess~\cite{smith1981,yehpatt1991}.
Runahead execution discards what it computed past a cache miss~\cite{mutlu2003runahead}.
A quantum state has no copy to fall back to, so \sys{} replaces the checkpoint with a checked recovery script and prices the recovery in the decision.

\textbf{Dynamic allocation.}
RushHour allocates resource states and ancilla space just in time, so a ready gate does not wait for statically reserved resources, and it applies each conditional correction after its decode~\cite{rushhour}.
Under speculation, such allocation can close the cycle of \secref{sec:cycle}, which static reservation prevents by holding a recovery's workspace from registration until the correction settles.

\textbf{Decoding and control.}
Decoders~\cite{skoric2023,tan2023,sparseblossom,unionfind,afs,lilliput,astrea,promatch,helios,microblossom,qecool,qulatis,ccdecoder,nisqplus,btwc} and real-time decoding~\cite{google2025,caune2024} shrink the wait but not the dependency~\cite{battistel2023}.
Syndrome data also drive leakage reduction and early aborts~\cite{gladiator,adabort}.
The QubiC and QICK controllers~\cite{qubic,qick} could host \sys{}'s decision unit and reservation manager.

\section{Conclusion}\label{sec:conclusion}

Reducing execution time matters in FTQC, but a fault-tolerant program stalls on its decoder, and speculative decoding still leaves its quantum operations waiting for confirmed results.
Quantum-side speculation removes that wait, but the system must recover from every misprediction on a state that cannot be copied.
In this work, we built \sys{}, which enables safe speculative execution on live quantum states.
Its compiler-assisted recovery contracts track how a misprediction propagates and give every stopping boundary a recovery script the compiler checks.
Its resource-aware runtime reserves every recovery workspace before it speculates on a correction and balances the benefit of speculating against the expected cost of recovery.

In simulation on top of SWIPER, \sys{} reduced quantum-side time against SWIPER alone by 9.81\% across all 88 configurations, by 16.62\% across the 50 where it speculates, and by 18.34\% across the 45 that allow concurrent predictions.
The reduction peaked at 31.1\% for QPE-5 at $d=21$.
At 38 configurations, \sys{} determined in training that speculation does not pay off, so it always waits and runs exactly as SWIPER alone does.

\begin{acks}
We used generative AI tools (Claude, GPT and Gemini) to run experiments and draft text.
We spent tons of hours manually checking all outcomes and take full responsibility for the work.
\end{acks}

\bibliographystyle{ACM-Reference-Format}
\bibliography{refs}

\end{document}